\documentstyle[12pt,aaspp4]{article}
\begin{document}

\title{Coalescing binary systems of compact objects: dynamics of angular 
momenta}
\author{Carlo Del Noce\altaffilmark{1}, Giovanni Preti\altaffilmark{2},
        and Fernando de Felice\altaffilmark{3}}
\altaffiltext{1}{delnoce@pd.infn.it}
\altaffiltext{2}{preti@pd.infn.it}
\altaffiltext{3}{defelice@pd.infn.it}
\affil{Dipartimento di Fisica ``G. Galilei'', Universit\`a di Padova, \\
       Via Marzolo 8, 35131, Padova, Italy, \\
       and INFN, Sezione di Padova}

\date{1997 September 22}

\begin{abstract}
The end state of a coalescing binary of compact objects depends strongly 
on the final total mass $M$ and angular momentum $J$.
Since gravitational radiation emission causes a slow evolution of 
the binary system through quasi-circular orbits down to the innermost
stable one, in this paper we examine the corresponding behavior of the ratio
$J/M^2$ which must be less than $1(G/c)$ or about $0.7(G/c)$ for the 
formation of a black hole or a neutron star respectively.
The results show cases for which, at the end of the inspiral phase, 
these conditions are not satisfied.
The inclusion of spin effects leads us to a study of precession 
equations valid also for the calculation of gravitational waveforms.
\end{abstract}

\begin{description}
\item[PACS number(s):] 04.25.Nx, 04.30.Db, 97.80.Fk, 97.60.Jd, 97.60.Lf.
\item[Subject headings:] binaries: close --- gravitation
\end{description}


\section{Introduction.}

Binary systems of compact objects like neutron stars or black holes are
promising sources of gravitational radiation (\cite{Tho}; \cite{Sch2});
in fact at coalescence they emit a great amount
of energy in the frequency range of LIGO and VIRGO detectors 
(\cite{Aea}).
At least a half of the observable stars come in binary or multiple systems
(\cite{Bat}) and a significant fraction of these may evolve to
neutron star or black hole binaries which would coalesce, due to
gravitational wave emission, in a time less than the age of the universe.
Moreover, recent estimates (\cite{Tho}; \cite{Sch1}; \cite{NPS};
\cite{Phi}; \cite{TY}; \cite{YSN}) 
make us confident that a few coalescences of compact binaries per year 
could be seen within a distance of about 100 Mpc.

In order to extract signal from noise, an extremely precise theoretical
prediction is needed for the gravitational waveform (\cite{Kid}).
Therefore much work was recently devoted to understand the evolution of the
system (\cite{Kid}; \cite{Dam}; \cite{LiWil}; \cite{KWW}; \cite{ACST};
\cite{BDI}, \cite{Bla}; \cite{WW}; and references therein).

An open question concerns the body coming out of coalescence.
It is well known that a stationary black hole of mass
$M$ and angular momentum $J$ must satisfy the condition
$J/M^2<1(G/c)$.
Also for a neutron star the ratio $cJ/(GM^2)$, hereafter defined as 
``Kerr parameter'', is less than about 0.7 (\cite{Cea}; \cite{FI};
\cite{SBGHa}, 1994b),
thus it is important to know how this ratio behaves during coalescence 
to see whether a black hole or a neutron star may actually result.

We also study the precession equations that describe the 
evolution of the bodies' spins.
Spin precession plays an important role in the physics of angular 
momentum, and its detailed knowledge is necessary for a consistent 
calculation of $J/M^2$.
Moreover spin precession is very important for the modulation of the 
gravitational wave (\cite{ACST}; \cite{Kid}; \cite{CF}).
Our results in this study apply directly to this field.

The coalescence of a binary system consists of two phases.
The first phase is a slow adiabatic inspiral of the orbits in which the
energy and angular momentum loss are assumed to be driven only by emission
of gravitational waves.
The second phase is highly dynamical: interactions between the bodies involve
their internal structure and full general relativistic effects must be
taken into account.

In this paper we consider only the first phase with the purpose of
determining the value of the Kerr parameter at the end
of it, which is the initial value of the Kerr parameter for the 
following phase of evolution in dynamical time.
This late-time evolution is not yet well understood, because 
its study is a formidable problem that can be 
attacked almost only by recourse to numerical methods.
It is essential for this task to know what the choice of 
initial values can be (\cite{Coo}), and
this paper is aimed as a help in understanding the state of the system 
in that critical moment of transition between the two phases.

We adopt the post-Newtonian formalism, which corresponds to a series
expansion
of all quantities in the parameter $v/c$ assumed to be small.
There are available in literature formulae for all physical quantities
pertaining to the physical system, which are valid through second 
post-Newtonian relative order,
i.e., including terms of order $(v/c)^4$ beyond the first nonzero one,
and including spin effects (\cite{Kid}).
This should be a pretty good approximation throughout the inspiral phase up
to the innermost stable circular orbit (ISCO) (\cite{JS}).

We shall neglect properties of the gravitating bodies connected with 
their internal structure except their spins: no correction terms for 
tidal interactions will be included since they would be equivalent to 
post-Newtonian corrections of higher order than what here wanted 
(\cite{LaWis}).
Again this approximation is less justified for compact bodies only 
when their separation becomes comparable to the ISCO radius.

Our calculations refer to quasi-circular orbits, for it has been proved
(\cite{Pet}; \cite{LiWil}) that all compact binaries have time enough
to evolve toward quasi-circular orbits before coalescence, except those that
are captured at near separations in highly eccentric orbits.

In \S \ref{ai} we consider the case of nonspinning bodies and 
examine the behavior of the corresponding Kerr parameter.
In \S \ref{aj} this analysis is extended to spinning bodies; in this 
case, however, it is essential to take into account the time evolution of 
the relative orientations of the orbital and intrinsic angular momenta.
Therefore \S \ref{ak} is focused on this topic.
Section \ref{al} then gives the results of the analysis of the Kerr 
parameter when spin precession effects are included.
Our conclusions are drawn in \S \ref{ao}.
We devote an Appendix to the study of the precession 
equations, due to their interest both for the problem at hand 
and more generally for the study of the gravitational waveform, 
and provide analytical solutions for particular cases.

\subsection{Conventions and Units.}   \label{ah}

Although we use measurement units in which the constant of light speed
in vacuo $c$ and the gravitational constant $G$ appear explicitly, 
we shall always make an effort to work with dimensionless 
physically meaningful variables.

We refer to a system of coordinates that satisfy the harmonic condition,
as usual in most related literature, and whose center of mass is fixed by the
same ``spin supplementary condition'' (SSC) used in (\cite{Kid}), that 
is the covariant SSC expressed by equation (A2a) in that reference.

We shall denote the ``Schwarzschild masses'' (\cite{Dam}) of the two 
bodies by $m_1$ and $m_2$, their sum by 
$m = m_1+m_2$ and the system's reduced mass 
by $\mu = m_1 m_2/(m_1+m_2)$, then set $\eta = \mu/m$.
Following Blanchet, Damour, \& Iyer (1995) we define $X_1 = m_1/m$, 
$X_2 = m_2/m$ and notice
that, since $X_1+X_2 = 1$ and $X_1 X_2=\eta$, $X_1$ and $X_2$ are functions 
of $\eta$ only: namely, $X_1=\frac{1}{2} \pm (\frac{1}{4}-\eta)^{1/2}$,
$X_2=\frac{1}{2} \mp (\frac{1}{4}-\eta)^{1/2}$.

Moreover let ${\bf x=x}_1-{\bf x}_2$ be the separation vector between the
two bodies in the chosen coordinate frame, $r = | {\bf x} |$,
${\bf v}= d{\bf x}/dt$, 
$\rho = (c^2/G) \; r/m$, $\gamma = (G/c^2) \; m/r=1/ \rho$, 
${\bf L}_N = \mu ({\bf x} \times {\bf v})$, 
$\hat{{\bf L}}_N ={\bf L}_N / |{\bf L}_N |$, ${\bf L} = {\bf 
L}_N$~+ post-Newtonian corrections, let $E$ be the system's Noetherian energy
(\cite{Dam}), $M = m+E/c^2$, and let ${\bf J}$ be the system's Noetherian 
total angular momentum (\cite{Dam}), $J = | {\bf J} |$.
We let ${\bf S}_i$ be the spin angular momentum of the $i$\/th body,
$S_i = | {\bf S}_i |$, $\hat{{\bf s}}_i = {\bf S}_i/S_i$ and
$\mbox{{\boldmath $\sigma$}}_i = (c/G)\; {\bf S}_i/m^2$ ($i=1,2$).
We write $\chi_i X_i^2 = \sigma_i = | \mbox{{\boldmath $\sigma$}}_i|$, so that
$\chi_i$ is the absolute value of the $i$\/th body's spin in units of
$(G/c) \; m_i^2$ or Kerr parameter, and ranges from 0 to 1 for a (Kerr) black
hole and from 0 to about 0.7 for a neutron star (\cite{FI}; 
\cite{SBGHa}, 1994b).
For purposes of later convenience in writing long equations we also define
$f = \hat{{\bf L}}_N\cdot\hat{{\bf s}}_1$, $g = \hat{{\bf L}}_N\cdot
\hat{{\bf s}}_2$, $h = \hat{{\bf s}}_1\cdot\hat{{\bf s}}_2$,
$V = \hat{{\bf L}}_N\cdot\hat{{\bf s}}_1\times\hat{{\bf s}}_2$.

\section{Nonspinning bodies.}   \label{ai}

As stated in the introduction, we assume here that the evolution 
of a close binary of compact objects can be approximately described 
by an adiabatic inspiral of pointlike masses through quasi-circular 
orbits, when the bodies' separation is large compared to their sizes 
(\cite{LiWil}).
The total mass of the system is given for both spinning and nonspinning
bodies and to the order relevant for our purposes 
by the following expression (\cite{WagW}; \cite{JS}):

\begin{equation}
M = m\left[1-\frac{\eta}{2}\gamma+\frac{\eta}{8}(7-\eta)\gamma^2+
O(\gamma^{5/2})\right] .
\label{a}    \end{equation}

In the case of nonspinning bodies ($S_1 = S_2 =0$) the general formula 
for ${\bf J}$ (\cite{Kid}) reduces to
\[
{\bf J}= \mu\left(G m r\right)^{1/2}\hat{{\bf L}}_N
\left[1+2\gamma+\frac{1}{2}(5-9\eta)\gamma^2+O(\gamma^{5/2})\right].
\]

Since we have assumed an adiabatic evolution of the system along 
the sequence of quasi-circular orbits, up to the innermost stable one 
at $r_{{\scriptsize \mbox{ISCO}}}\approx 6Gm/c^2$, the actual
expressions of the mass and angular momentum loss rate due to gravitational
radiation emission are not essential to evaluate the behavior of 
$M$ and ${\bf J}$. 
We can then study directly the behavior of the Kerr parameter.
We obtain:
\begin{equation}
\frac {c}{G}\frac{J}{M^2}=\eta\rho^{1/2}\left[1+(2+\eta)\gamma+
\frac{1}{4}\left(10 - 17 \eta +4 \eta^2 \right)\gamma^2+
O(\gamma^{5/2})\right] .                        \label{c}
\end{equation}

Since $\eta$ ranges between 0 and $\frac{1}{4}$, the latter
corresponding to a system of two bodies with equal masses,
any deviation from equipartition of masses in the binary system
would decrease the Kerr parameter.

Figure 1 is a contour plot of $cJ/(GM^2)$ in the
($\rho$, $\eta$)-plane.

The approximation $S_1=S_2=0$ is good for the description of systems
consisting of two slowly spinning neutron stars, since the common values of
their Kerr parameters are less or much less than 
$10^{-2}$ (\cite{dFY}).
We notice that any process of slow 
energy and angular momentum dissipation that keeps the system in 
quasi-circular orbits decreases 
the Kerr parameter below unity just at the limit of validity of our
approach ($\rho \approx 6$).
At this separation the conditions for the formation of a 
stationary rotating neutron star are hardly met; first because 
$J/M^2 < 0.7(G/c)$ holds only if $\eta$ is less than about 0.2, which 
corresponds to a more massive body about 2.6 times as large as the other,
and this is not the case for a system consisting initially of neutron 
stars, second because the total mass $M$ must be less than the 
limiting mass of neutron stars.
Thus we conclude from an inspection of Figure 1 that, since the most 
probable case for a neutron star binary is that they have nearly equal 
masses, the outcome of a final coalescence 
could be either a fast rotating black hole or a highly rotationally excited 
neutron star.

\subsection{Estimate of accuracy.}       \label{aq}

It is not at all straightforward to estimate the accuracy of 
equation (\ref{c}) for all choices of $\eta$ and all values of $\rho$.
In order to perform a rigorous estimate we should either know something 
about the convergence of the post-Newtonian series (eq.\ [\ref{c}]) 
or be able to solve the problem exactly.
Instead our information on the post-Newtonian series is rather poor.
We neither have an upper bound for the series remainder nor know 
anything about the first neglected terms.
In fact we exploited all the information in our hands---that is the 
post-Newtonian terms known up to date---for our calculations.
Moreover there are hints of slow convergence of 
post-Newtonian series, reported by some recent studies (\cite{Poi};
\cite{SLPW}).
If we assume that the largest error source in truncating the series is 
given by the first neglected term, we can take the term of order 
$\gamma^{5/2}$ in brackets in equation (\ref{c}) as the relative error,
or better, as its order of magnitude.
Moreover, we shall assume that the coefficient of the term 
$\gamma^{5/2}$ is 1, a simplification possible because
our units are such that all coefficients in the post-Newtonian 
series are of order 1.
Some values for the relative error thus estimated are given below:
\[ \begin{array}{lll}
\mbox{error} \approx 3 \cdot 10^{-8} , & ~~~~~~~~ & \mbox{at~} \rho
= 10^{3} , \\
\mbox{error} \approx 10^{-5} , & ~~~~~~~~ & \mbox{at~} \rho
= 10^{2} , \\
\mbox{error} \approx 0.11 , & ~~~~~~~~ & \mbox{at~} \rho
= 6 . \end{array}
\]

Our confidence in the above estimate may be strengthened by a comparison 
with the case of a test mass in circular orbit around a Schwarzschild 
black hole.
This problem of a test mass and a Schwarzschild black hole is well known 
and exactly solved.
Actually it provides almost a benchmark for all post-Newtonian 
calculations, since agreement in the limit of $\eta$ going to zero gives 
a necessary condition for their validity.
We might also conjecture (\cite{KWW}) that this comparison 
gives a pessimistic estimate of the error, since the post-Newtonian 
series is poorly convergent in the Schwarzschild case.
The formula analogous to equation (\ref{c}) for a test mass in circular orbit 
around a black hole reads
\[
\frac {c}{G}\frac{J}{M^2}=\eta\rho^{1/2}\frac{\sqrt{(1+\gamma)^2/
(1-2\gamma)}}{\left\{ 1+\eta\sqrt{(1-\gamma)^2/[(1+\gamma)(1-2\gamma)]}
\; \right\}^2} .
\]
Since both equation (\ref{c}) and the above equation are 0 when $\eta$ is 0, 
we actually shall divide them by $\eta$, before any comparison.
Thus the second post-Newtonian development of the above formula 
coincides with equation (\ref{c}) at $\eta=0$.
The difference between the two formulae---which is a rigorous estimate of 
the post-Newtonian terms neglected in equation (\ref{c}) in the  limit 
$\eta=0$---is very small, being less that $10^{-3}$ for 
$\rho \gtrsim 28.3$, and showing a sudden increase for 
decreasing $\rho$ only at $\rho \lesssim 16$ (due to 
Schwarzschild behavior) but keeping anyway 
below $6.4\times 10^{-2}$ for $\rho \geq 6$.

\section{Spinning bodies.}          \label{aj}

Most celestial bodies have spin; therefore
it is interesting to study how intrinsic rotation would affect the
evolution of the Kerr parameter for coalescing binaries.

As in the previous case, we shall confine ourselves to
post-Newtonian corrections of order $\gamma^2$ at most.
Since spin-orbit and spin-spin interaction terms will affect the
total mass only at order equal to or higher than $\gamma^{5/2}$,
inclusion of spin will not change the equation (\ref{a}) for the total mass.
On the other hand, the post-Newtonian corrections to
the angular momentum will include spin effects already at
relative order $\gamma^{3/2}$.
With the assumption of a slow precession of the orbital angular momentum and
spins around the total angular momentum, which is valid as long as
$\gamma \ll 1$ (\cite{Kid}), we may take the 
equation for ${\bf J}$ averaged over one orbit (\cite{Kid}) 
and rewrite it with our notations as
\begin{equation}
\begin{array}{rl}
{\bf J} = & {\displaystyle \mu (G m r)^{1/2}\hat{{\bf L}}_N
\left( 1 +2\gamma-\left[(\hat{{\bf L}}_N\cdot\mbox{{\boldmath $\sigma$}}_1)
\left(2+\frac{7}{4}\frac{m_2}{m_1}\right)+(\hat{{\bf L}}_N\cdot
\mbox{{\boldmath $\sigma$}}_2)\left(2+\frac{7}{4}
\frac{m_1}{m_2}\right)\right]\gamma^{3/2}+ \right. }\\  { } & { } \\
&{\displaystyle \left. +\left\{ \frac{1}{2}(5-9\eta)-\frac{3}{4\eta}
[(\mbox{{\boldmath $\sigma$}}_1\cdot\mbox{{\boldmath $\sigma$}}_2)-3
(\hat{{\bf L}}_N\cdot\mbox{{\boldmath $\sigma$}}_1)
(\hat{{\bf L}}_N\cdot\mbox{{\boldmath $\sigma$}}_2)]\right\} \gamma^2+
O(\gamma^{5/2})\right) + } \\   { } & { } \\
& {\displaystyle +{\bf S}_1\left[1-\frac{1}{4}(3\eta+X_2)\gamma+O(\gamma^2)
\right]+{\bf S}_2\left[1-\frac{1}{4}(3\eta+X_1)\gamma+O(\gamma^2) \right] .}
\end{array}
\label{d}
\end{equation}

We now calculate the square power of $J/M^2$ from equations 
(\ref{d}) and (\ref{a}) and obtain
\begin{equation}
{\displaystyle \left(\frac{c}{G}\frac{J}{M^2}\right)^2 =
\rho \left[A_0+A_1\gamma^{1/2}+A_2\gamma+A_3\gamma^{3/2}+A_4\gamma^2+
O(\gamma^{5/2})\right]} ,              \label{e}
\end{equation}
where:
\begin{equation}    \left\{    \begin{array}{rl}
A_0=&\eta^2, \\ 
A_1=&2\eta[\hat{{\bf L}}_N\cdot(\mbox{{\boldmath $\sigma$}}_1+
\mbox{{\boldmath $\sigma$}}_2)] , \\ 
A_2=&2\eta^2(2+\eta)+\sigma_1^2+\sigma_2^2+2(\mbox{{\boldmath $\sigma$}}_1\cdot
\mbox{{\boldmath $\sigma$}}_2) , \\  
A_3=&2\eta[(\hat{{\bf L}}_N\cdot\mbox{{\boldmath $\sigma$}}_1)(2X_1+\eta)+
  (\hat{{\bf L}}_N\cdot\mbox{{\boldmath $\sigma$}}_2)(2X_2+\eta)] , \label{f}\\
A_4=&\frac{3}{2}\eta^2(6-3\eta+2\eta^2)-\frac{1}{2}(1+\eta)
(\mbox{{\boldmath $\sigma$}}_1\cdot\mbox{{\boldmath $\sigma$}}_2)-
\frac{7}{2}(1-\eta)(\hat{{\bf L}}_N\cdot\mbox{{\boldmath $\sigma$}}_1)
(\hat{{\bf L}}_N\cdot\mbox{{\boldmath $\sigma$}}_2) + \\
& -{\displaystyle \frac{X_2}{2}[X_2\sigma_1^2+(\hat{{\bf L}}_N\cdot
\mbox{{\boldmath $\sigma$}}_1)^2(8-X_2)]-\frac{X_1}{2}[X_1\sigma_2^2+
(\hat{{\bf L}}_N\cdot\mbox{{\boldmath $\sigma$}}_2)^2(8-X_1)] } .
\end{array}   \right.        \label{ar} \end{equation}

We shall deal with equation (\ref{e}) rather than try to deduce a post-Newtonian 
expression for $J/M^2$ as a truncated series because deciding
which term in equation (\ref{e}) is the 
leading one depends on the choice of the parameters $\eta$, 
$\chi_1$ and $\chi_2$ and on the relative orientations of $\hat{{\bf 
L}}_N$, $\mbox{{\boldmath $\sigma$}}_1$ and $\mbox{{\boldmath $\sigma$}}_2$. 
As a matter of fact, if we focus on the first few terms of equation 
(\ref{e}), we can write 
\[
{\displaystyle \left(\frac{c}{G}\frac{J}{M^2}\right)^2 =
[\eta\rho^{1/2}(1+2\gamma)\hat{{\bf L}}_N+\mbox{{\boldmath $\sigma$}}_1+
\mbox{{\boldmath $\sigma$}}_2]^2+2\eta^3+O(\gamma^{1/2}) }
\]
and notice that the Kerr parameter for spinning bodies 
depends mainly on the variables of the vector 
$\eta\rho^{1/2}(1+2\gamma)\hat{{\bf L}}_N+\mbox{{\boldmath $\sigma$}}_1+
\mbox{{\boldmath $\sigma$}}_2$ (that is, the first post-Newtonian form of
${\bf J}$ in units of $Gm^2/c$) which determine the relative weight of 
the leading terms in equation (\ref{e}).
For example, if $\chi_1=\chi_2=1$ and $\eta=\frac{1}{7}$ (which corresponds to a 
more massive body about 5 times as large as the other), 
$A_2\gamma$ is of the same order of magnitude as $A_0$ already at values of
$\rho \approx 50$ and greater at shorter separations.
Let us remark that this does not imply that the second post-Newtonian 
formula (eq.\ [\ref{e}]) is no longer valid at $\rho \approx 50$, because the 
parameter of development of the post-Newtonian series is $\gamma$ 
which does not depend on $\eta$, $\sigma_1$ and $\sigma_2$.
We just point out that calculating $J/M^2$ needs care.

As for the accuracy of equation (\ref{e}), considerations quite similar 
to those stated in \S \ref{aq} about the post-Newtonian series 
(eq.\ [\ref{c}]) apply.
A thorough discussion should take into account the complicated form of 
the coefficients of the post-Newtonian series (eq.\ [\ref{e}]; see eq.\ 
[\ref{ar}] and the unknown similar expressions for the neglected terms).
Anyway, since such coefficients cannot be much larger than unity, the 
conclusions we reach are substantially equivalent.

As expected and already noticed, equation (\ref{e}) depends on the 
relative orientations of the orbital and intrinsic angular momenta of 
the coalescing bodies, namely, on the quantities 
$\hat{{\bf L}}_N\cdot\mbox{{\boldmath $\sigma$}}_1$, 
$\hat{{\bf L}}_N\cdot\mbox{{\boldmath $\sigma$}}_2$, and
$\mbox{{\boldmath $\sigma$}}_1\cdot\mbox{{\boldmath $\sigma$}}_2$, which
are functions of the coordinate time $t$ through $r$.
Such functions were considered already by Cutler \& Flanagan (1994) 
because of 
their importance to the secular growth of the gravitational wave phase.
We performed a detailed analysis of these functions, finding results 
that we shall report in Appendix, since they have more general 
application than just in our problem.
A related and interesting topic is also the evolution of the absolute 
orientations of $\hat{{\bf L}}_N$, $\mbox{{\boldmath $\sigma$}}_1$, 
$\mbox{{\boldmath $\sigma$}}_2$, since the observed gravitational 
waveform depends on the angle between the orbital plane---almost 
orthogonal to $\hat{{\bf L}}_N$ with post-Newtonian corrections along 
$\mbox{{\boldmath $\sigma$}}_1$ and $\mbox{{\boldmath $\sigma$}}_2$---and 
the wave propagation direction.
This question is already fully examined in literature (\cite{Kid}; 
\cite{ACST}), therefore we shall not consider it.

\subsection{Effects of spin precession.}   \label{ak}

Here we shall deduce the equations to solve for determining the 
coefficients of equation (\ref{ar}) under the following two 
assumptions: first, that the spins precess while keeping constant 
absolute values and that the gravitational radiation emission only 
affects the orbital angular momentum ${\bf L}$ (\cite{ACST}; 
\cite{Kid}); 
second, that the precession frequency is much less than the orbital 
frequency.
As analyzed by Kidder (1995) the ratio of the precession frequency to 
the orbital frequency goes as $\gamma$.

We are interested in the dependence on $r$ of the scalar products 
between any pair of the three vectors $\hat{{\bf L}}_N$, 
$\mbox{{\boldmath $\sigma$}}_1$, $\mbox{{\boldmath $\sigma$}}_2$.
We put together the equations for the orbital angular momentum, 
separation evolution, and spin precession (\cite{Kid}) averaged 
over one orbit to lowest order and have
\[      {\bf L}_N = \mu (G m r)^{1/2} \hat{{\bf L}}_N,
~~~~~~~~~~~~~~~~~~~~~~~~~~~~
\frac{d r}{d t} = -\frac{64}{5}c\eta\gamma^3,    \]
\begin{equation}    \left\{     \begin{array}{rl}
{\displaystyle \frac{d {\bf L}_N}{d t}} = & {\displaystyle \frac{G}{2c^2r^3}
\left\{\left[\left(4+3\frac{m_2}{m_1}\right){\bf S}_1+\left(4+3\frac{m_1}{m_2}
\right){\bf S}_2\right]\times {\bf L}_N + \right.} \\
& {\displaystyle \left. -3[(\hat{{\bf L}}_N\cdot{\bf S}_2){\bf S}_1+
(\hat{{\bf L}}_N\cdot{\bf S}_1){\bf S}_2]\times\hat{{\bf L}}_N\right\}-
\frac{32}{5}\frac{c^3}{G}\frac{\eta}{m}\gamma^4{\bf L}_N } , \\  {} & {} \\
{\displaystyle \frac{d {\bf S}_1}{d t}} = & {\displaystyle \frac{G}{2c^2r^3}
\left[\left(4+3\frac{m_2}{m_1}\right){\bf L}_N+{\bf S}_2-3(\hat{{\bf L}}_N\cdot
{\bf S}_2)\hat{{\bf L}}_N\right]\times{\bf S}_1 }   , \\  {} & {} \\
{\displaystyle \frac{d {\bf S}_2}{d t}} = &{\displaystyle \frac{G}{2c^2r^3}
\left[\left(4+3\frac{m_1}{m_2}\right){\bf L}_N+{\bf S}_1-3(\hat{{\bf L}}_N\cdot
{\bf S}_1)\hat{{\bf L}}_N\right]\times{\bf S}_2} .
\end{array} \right.    \label{i}      \end{equation}

Combining the above equations and rewriting the result in terms 
of dimensionless quantities we obtain:
\begin{equation}     \left\{         \begin{array}{rl}
{\displaystyle \frac{d}{d \rho}(\hat{{\bf L}}_N\cdot
\mbox{{\boldmath $\sigma$}}_1)} = &
{\displaystyle -\frac{15}{128\eta}\left[\frac{1}{X_2}-\frac{(\hat{{\bf L}}_N
\cdot
\mbox{{\boldmath $\sigma$}}_1)}{\eta\rho^{1/2}}\right](\hat{{\bf L}}_N\cdot
\mbox{{\boldmath $\sigma$}}_1\times\mbox{{\boldmath $\sigma$}}_2) } , \\ {}&{} \\
{\displaystyle \frac{d}{d \rho}(\hat{{\bf L}}_N\cdot
\mbox{{\boldmath $\sigma$}}_2)} = &
{\displaystyle \frac{15}{128\eta}\left[\frac{1}{X_1}-\frac{(\hat{{\bf L}}_N
\cdot
\mbox{{\boldmath $\sigma$}}_2)}{\eta\rho^{1/2}}\right](\hat{{\bf L}}_N\cdot
\mbox{{\boldmath $\sigma$}}_1\times\mbox{{\boldmath $\sigma$}}_2) } , \\ {}&{} \\
{\displaystyle \frac{d}{d \rho}(\mbox{{\boldmath $\sigma$}}_1\cdot
\mbox{{\boldmath $\sigma$}}_2)} = &
{\displaystyle -\frac{15}{128\eta}\left[\left(\frac{m_2}{m_1}-\frac{m_1}{m_2}
\right)\eta
\rho^{1/2}+(\hat{{\bf L}}_N\cdot\mbox{{\boldmath $\sigma$}}_1)+ \right. } 
\\ {}&{} \\ 
& {\displaystyle \left. -(\hat{{\bf L}}_N\cdot
\mbox{{\boldmath $\sigma$}}_2)\right](\hat{{\bf L}}_N\cdot
\mbox{{\boldmath $\sigma$}}_1\times\mbox{{\boldmath $\sigma$}}_2) } .
\end{array}       \right.    \label{k}    \end{equation}

Expanding the quantity 
$[\hat{{\bf L}}_N\times(\mbox{{\boldmath $\sigma$}}_1\times
\mbox{{\boldmath $\sigma$}}_2)]^2$ first according
to the general rule for ${\bf a}\times({\bf b}\times{\bf c})$, then according
to that for $({\bf a}\times{\bf b})\cdot({\bf c}\times{\bf d})$, 
we get an equation to close system (\ref{k}):
\begin{equation}     \begin{array}{rl}
(\hat{{\bf L}}_N\cdot\mbox{{\boldmath $\sigma$}}_1\times
\mbox{{\boldmath $\sigma$}}_2)^2 = &
\sigma_1^2\sigma_2^2-(\mbox{{\boldmath $\sigma$}}_1\cdot
\mbox{{\boldmath $\sigma$}}_2)^2-\sigma_1^2
(\hat{{\bf L}}_N\cdot\mbox{{\boldmath $\sigma$}}_2)^2-\sigma_2^2
(\hat{{\bf L}}_N\cdot\mbox{{\boldmath $\sigma$}}_1)^2+ \\
& +2(\mbox{{\boldmath $\sigma$}}_1\cdot\mbox{{\boldmath $\sigma$}}_2)
(\hat{{\bf L}}_N\cdot\mbox{{\boldmath $\sigma$}}_1)(\hat{{\bf L}}_N\cdot
\mbox{{\boldmath $\sigma$}}_2) .
\end{array}      \label{p}     \end{equation}

The system of equations (\ref{k}) contains nonlinear differential equations 
whose solutions have an oscillatory behavior.
In order to avoid that a numerical integration stops when 
$\hat{{\bf L}}_N\cdot \mbox{{\boldmath $\sigma$}}_1\times\mbox{{\boldmath 
$\sigma$}}_2$ first reaches 0, we also differentiate equation (\ref{p}) 
and substitute it with the following:
\begin{equation}        \begin{array}{rl}
{\displaystyle \frac{d}{d\rho} (\hat{{\bf L}}_N\cdot
\mbox{{\boldmath $\sigma$}}_1\times\mbox{{\boldmath $\sigma$}}_2) } & =  
{\displaystyle -\frac{15}{128\eta} \left\{ (X_1-X_2)[
(\mbox{{\boldmath $\sigma$}}_1\cdot\mbox{{\boldmath $\sigma$}}_2)-
(\hat{{\bf L}}_N\cdot\mbox{{\boldmath $\sigma$}}_1)
(\hat{{\bf L}}_N\cdot\mbox{{\boldmath $\sigma$}}_2)]\rho^{1/2}+ \right. } \\ 
{}&{} \\ 
& {\displaystyle -(\hat{{\bf L}}_N\cdot\mbox{{\boldmath $\sigma$}}_1)
(\mbox{{\boldmath $\sigma$}}_1\cdot\mbox{{\boldmath $\sigma$}}_2)
\frac{1+X_1}{X_1}+ (\hat{{\bf L}}_N\cdot\mbox{{\boldmath $\sigma$}}_2)
(\mbox{{\boldmath $\sigma$}}_1\cdot\mbox{{\boldmath $\sigma$}}_2)
\frac{1+X_2}{X_2}+}  \\{}&{} \\
& {\displaystyle + (\hat{{\bf L}}_N\cdot\mbox{{\boldmath $\sigma$}}_1)
(\hat{{\bf L}}_N\cdot\mbox{{\boldmath $\sigma$}}_2)
(\hat{{\bf L}}_N\cdot\mbox{{\boldmath $\sigma$}}_1-
\hat{{\bf L}}_N\cdot\mbox{{\boldmath $\sigma$}}_2)
-\frac{\sigma_2^2}{X_2}(\hat{{\bf L}}_N\cdot\mbox{{\boldmath $\sigma$}}_1)+}
\\ {}&{} \\
& {\displaystyle 
+\frac{\sigma_1^2}{X_1}(\hat{{\bf L}}_N\cdot\mbox{{\boldmath $\sigma$}}_2)+ 
\left. \frac{\sigma_2^2
(\hat{{\bf L}}_N\cdot\mbox{{\boldmath $\sigma$}}_1)^2-\sigma_1^2
(\hat{{\bf L}}_N\cdot\mbox{{\boldmath $\sigma$}}_2)^2}
{\eta \rho^{1/2}} \right\} } .
\end{array}               \label{o}            \end{equation}

We have solved the system of equations (\ref{k}), (\ref{o}) 
numerically for a few choices of the parameters 
$\chi_1$, $\chi_2$, $X_1/X_2$ and of the initial values.
Typical behaviors of the solutions to the above equations will be 
shown in the figures that illustrate the Appendix.
We have judged the analysis of the system of equations (\ref{k}), (\ref{o}) 
explained in Appendix important for understanding the role of spin precession 
during coalescence, since in general the behavior of the relative 
orientations among $\hat{{\bf L}}_N$, $\mbox{{\boldmath $\sigma$}}_1$, 
and $\mbox{{\boldmath $\sigma$}}_2$ is not simply predictable: for 
example even the average values of the oscillating functions do not 
keep fixed but evolve, as can be seen in Figure 2---an example in 
which this effect is dramatic.

\subsection{Results.}    \label{al}

Once the evolution of
$\hat{{\bf L}}_N\cdot\mbox{{\boldmath $\sigma$}}_1$, 
$\hat{{\bf L}}_N\cdot\mbox{{\boldmath $\sigma$}}_2$, 
$\mbox{{\boldmath $\sigma$}}_1\cdot\mbox{{\boldmath $\sigma$}}_2$ is 
known, we can deduce the behavior of the Kerr parameter from equations
(\ref{e}) and (\ref{f}).
Plots of this quantity for $\chi_1=\chi_2=1$ and different mass ratios 
are shown in Figure 3 ({\em solid lines}) and compared to the
corresponding spinless case ({\em dashed lines}).

If the intrinsic spins of the bodies are not very small and if they are 
directed with a positive component along the orbital angular momentum, 
then the value of $J/M^2$ remains 
larger than $1(G/c)$ over the entire inspiral phase (Fig.\ 3a).
When one body is much more massive than the other and its spin is 
initially positively oriented with the orbital angular momentum, then it 
contributes to keep the Kerr parameter larger than 1 even if the 
companion's spin is negatively oriented, as shown in Fig.\ 3b.
When both spins are negatively oriented with respect to the orbital angular 
momentum, then the Kerr parameter is less than in the corresponding 
spinless case, at least initially.
In fact, at the end of the inspiral, the orbital angular momentum has 
been almost completely radiated away and only the spin is left over, 
giving a positive contribution to $J/M^2$ irrespective of its 
direction.
For intermediate separations of the two bodies, it may happen that the 
spin corrections to $J/M^2$ at different post-Newtonian orders add up 
to 0, as in the case of Figure 3c (notice the crossing of the two 
curves).
It may also happen that total spin and orbital angular momentum have 
vanishing sum, as shown in Figure 3d (notice the solid curve approaching 
zero); if this phenomenon occurs before final coalescence, the 
evolution of the angular momenta is known as transitional precession 
(\cite{ACST}).
In this last case the value of $\rho$ at which $J/M^2$ approaches zero 
varies with the system's mass ratio and initial directions of the 
angular momenta.
Thus there are particular choices of these values 
such that the Kerr parameter decreases almost to zero at the 
end of the inspiral phase; an example is shown in Figure 4  
(the apparent flattening of the curve at $\rho\approx 6$ is 
uncertain since the approximation there reaches its limit of validity).
The need for special and uncommon choices (see discussion by Apostolatos 
et al.\ (1994), \S IV.D.1) of the parameters in order to obtain a final vanishing 
total angular momentum makes us think that the formation of a 
Schwarzschild black hole, as result of coalescence of a compact binary, 
is a rare event.

The dynamics of inspiralling bodies and of their angular momenta, as 
described by the above formulae, is exactly the same even if the bodies 
are massive black holes.
We can infer from our analysis that the end of the 
inspiral phase of coalescence of two black holes in the center of an active 
galactic nucleus is most likely characterized by conditions which are 
compatible with the formation of a fast rotating black hole.

Since $J/M^2$ has a complicated dependence on 
$\hat{{\bf L}}_N\cdot\hat{{\bf s}}_1$, $\hat{{\bf L}}_N\cdot
\hat{{\bf s}}_2$, $\hat{{\bf s}}_1\cdot\hat{{\bf s}}_2$,
and the parameters $\eta$, $\chi_1$, $\chi_2$, we computed the 
probability density function of $cJ/(GM^2)$ at $\rho=6$ numerically 
(by the Monte Carlo method) for some plausible choices of parameters 
corresponding to neutron star binary systems.
The probabilities of $m_1$ and $m_2$ were assumed uniformly distributed 
over an interval ranging from 1.2 to 1.6 (the scale does not matter) and 
the orientations of $\hat{{\bf L}}_N$, $\hat{{\bf s}}_1$, 
$\hat{{\bf s}}_2$ were taken at random, with uniform probability, among 
all directions of space.
The parameters $\chi_1$ and $\chi_2$ were also taken at random but with a probability 
distribution uniform in $\log \chi$ and ranging from $7.96 \times 
10^{-5}$ (corresponding to the period of PSR J1951+1123, that is the 
largest known pulsar period; \cite{CN}) to 0.7 (corresponding 
to about the highest value compatible with rotational stability).
The result is the extremely peaked function plotted in Figure 5a.
It has a mean equal to 0.86 with standard deviation $2.94 \times 10^{-2}$, 
coefficient of skewness +7.67 and coefficient of excess (kurtosis)
equal to +6.93.
A normal distribution with same mean and deviation is plotted (without 
tails) for comparison ({\em dashed line}).

In Figure 5b the same distribution curve as in Figure 5a is compared to 
the probability distribution of $J/M^2$ in the spinless case (again 
numerically calculated by the Monte Carlo method and with the same 
assumed mass distribution function as before).
This latter curve, dashed in figure, is even more peaked than 
the former, and it reaches its maximum at the value that equation
(\ref{c}) takes on for $\rho=6$ and $\eta=\frac{1}{4}$, that is $J/M^2 
\approx 0.8675 (G/c)$.

\section{Conclusions.}            \label{ao}

Neglecting finite size effects we examine the evolution of the Kerr 
parameter all the way up to the state in which the separation between 
the bodies is $6Gm/c^2$ corresponding about to the onset of the 
dynamical instability of circular orbits.
We found cases (for example, those shown in Figs.\ 5a and 5b) in which 
it remains larger than 1 (another instance was 
also found in the numerical study by Wilson, Mathews, \& Marronetti 
1996).
This implies that the formation of a stationary black hole is possible 
in those cases only if $J/M^2$ is decreased efficiently in the final 
dynamical phases of coalescence.
As a matter of fact we cannot infer anything on the nature of 
the coalesced body in such cases.
An argument like the one in the discussion by Cook (1994), 
concluding for a final Kerr parameter less than 1 without 
going through the full solution of the evolution problem, simply is not 
valid: first, because it would reach the same conclusion 
irrespectively 
of the value of $J/M^2$ at the beginning of the dynamical 
phase (eq.\ [16] in that paper always gives $J/M^2 \leq 1$); 
second, because it is based on the ``a priori'' not 
justified assumption of having a Kerr black hole at the end of the 
coalescence, and this already by itself implies that $J/M^2 \leq 1$.

Similarly the formation of a stationary neutron star seems highly 
unlikely in the absence of a powerfully dissipative mechanism, since we 
found that in all cases except a few very special ones $J/M^2$ is still 
greater than $0.7 (G/c)$ at the end of the inspiral phase.
Anyway, when the final coalesced object is a black hole, our results 
give us confidence that it is rapidly rotating (\cite{Coo}).

In the course of studying our main object, we had to deal with the 
problem of precessing spins.
We found analytical solutions to the precession equations which we 
reported since they have a wider application than just to our purpose:
that is, to the calculation of the gravitational wave modulation due to spin 
precession, as fully explained by Cutler \& Flanagan (1994).

\acknowledgments{
This work was partially supported by the
Italian Space Agency (ASI), the Ministero della Universit\`a e della
Ricerca Scientifica e Tecnologica (MURST) of Italy and by the GNFM
of the Italian Consiglio Nazionale delle Ricerche (CNR).
We would also thank the scientific editor, B. Haisch, and the 
anonymous referee for suggestions and positive criticisms that led to 
improvements of our paper.
}

\appendix
\section{Analytical solutions of the 
          precession equations.}  \label{am}

A better insight into the behavior of the solutions to the system of 
differential equations (\ref{k}), (\ref{p}) is ensured by an 
analytical study of them.
Because of the complexity of the problem, we have been able to obtain 
only approximate solutions for some cases.
We have judged this analytical study useful because it applies also to 
the calculation of the gravitational waveform (\cite{CF}).

\subsection{Preliminary considerations.}

By elimination of $\hat{{\bf L}}_N\cdot\mbox{{\boldmath $\sigma$}}_1\times
\mbox{{\boldmath $\sigma$}}_2$ from equation (\ref{k}) we find two relations that 
do not involve $\hat{{\bf L}}_N\cdot\mbox{{\boldmath $\sigma$}}_1\times
\mbox{{\boldmath $\sigma$}}_2$, namely,
\begin{equation}     \left\{  \begin{array}{l}     
{\displaystyle \rho^{-1/2} \frac{d}{d \rho}(\mbox{{\boldmath $\sigma$}}_1
\cdot\mbox{{\boldmath $\sigma$}}_2)
+\eta\frac{d}{d \rho}[\hat{{\bf L}}_N\cdot(\mbox{{\boldmath $\sigma$}}_1+
\mbox{{\boldmath $\sigma$}}_2)]=0 } , \\   {} \\
{\displaystyle \rho^{-1/2} \frac{d}{d \rho}[(\hat{{\bf L}}_N\cdot
\mbox{{\boldmath $\sigma$}}_1)
(\hat{{\bf L}}_N\cdot\mbox{{\boldmath $\sigma$}}_2)] =
\frac{d}{d \rho}[X_2(\hat{{\bf L}}_N\cdot\mbox{{\boldmath $\sigma$}}_1)+
X_1(\hat{{\bf L}}_N\cdot\mbox{{\boldmath $\sigma$}}_2)] } .
\end{array}         \right.         \label{l}       \end{equation}

Equations (\ref{l}) are of the form $d\psi/d\rho+\rho^{-1/2}
d\varphi/d\rho = 0$, hence an integration by parts yields 
$\psi+\rho^{-1/2}\varphi+\frac{1}{2}
\int^{\rho}\rho^{-3/2}\varphi(\rho)\,d\rho = 0$, and
since $\varphi$ is in all cases at most of the order of unity, 
then for $\rho \gg 1$, this last equation leads to ``approximate''
first integrals in the following form:
\[         \psi = \mbox{constant} + O(\rho^{-1/2}) .                   \]

In order to deal with quantities which have manifestly the same
order of magnitude, we find it convenient to rewrite the system of 
equations (\ref{k})--(\ref{p}) in terms of the variables $f$, $g$, $h$, 
and $V$:
\begin{eqnarray}
\frac{df}{d \rho}  & =& {\displaystyle 
-\frac{15}{128}\left(\frac{\chi_2}{X_1}
-\chi_1\chi_2f\rho^{-1/2}\right)V , }    \label{w} \\ 
{} & {} & {}    \nonumber \\
\frac{dg}{d \rho} & =& {\displaystyle 
\frac{15}{128}\left(\frac{\chi_1}{X_2}
-\chi_1\chi_2g\rho^{-1/2}\right)V , }    \label{x}  \\
{} & {} & {}    \nonumber \\
\frac{dh}{d \rho}&  = & {\displaystyle 
-\frac{15}{128}\left[\left(\frac{X_2}{X_1}
-\frac{X_1}{X_2}\right)\rho^{1/2}+\chi_1\frac{X_1}{X_2}f-\chi_2\frac{X_2}{X_1}
g\right]V , } \label{n}        \\
{} & {} & {}    \nonumber \\
V^2&=&1-f^2-g^2-h^2+2fgh .                        \label{y}   
\end{eqnarray}

Let us denote the initial values of $f$, $g$, $h$, and $\rho$ by $f_0$, 
$g_0$, $h_0$, and $\rho_0$ respectively, and the initial value of
$\hat{{\bf L}}_N\cdot\hat{{\bf s}}_1\times\hat{{\bf s}}_2$,
with its sign, by $V_0$.
Moreover we shall denote the initial value of the right-hand side of equation
(\ref{o}) by $\sigma_1 \sigma_2 V'_0$.

\subsection{One spinning body.}

In the very special case where $S_1=0$ or $S_2=0$ ``exactly'',
we have
\[
\hat{{\bf L}}_N\cdot\hat{{\bf s}}_1= \mbox{constant ~~~~~~~(if $S_2=0$) } , 
\] 
trivially from equation (\ref{i}).

\subsection{Equal masses.}  \label{u}

Let $X_1=X_2$ and thus $\eta=1/4$.
Then the first 
term in the right-hand side of equation (\ref{n}) vanishes.
Moreover from equations (\ref{l}) we obtain the following first integral 
``exactly'' (in the sense that it is a necessary consequence of eq.\ 
[\ref{l}], with no need for any further approximation):
\begin{equation}   
h+\frac{1}{2}fg= \mbox{constant}  .
\label{q} \end{equation}
On the other hand, if $\rho \gg 1$, we can neglect the terms containing 
$\rho^{-1/2}$ in the right-hand sides of equations (\ref{w}) and (\ref{x}) 
and take either equation in (\ref{l}) (but not both) to give an 
``approximate'' first integral as (cf.\ eq.\ 46 in \cite{ACST}) 
\begin{equation}             
\chi_1f+\chi_2g = \mbox{constant}  .        \label{aa}
\end{equation}

Summing up, we can approximate the system of equations (\ref{w})--(\ref{y}) 
with the following autonomous differential system:
\begin{equation}
\left\{
\begin{array}{l}
{\displaystyle \frac{df}{d \rho} = -\frac{15}{64}\chi_2V } , \\
{\displaystyle \frac{dg}{d \rho} = \frac{15}{64}\chi_1V } , \\  
{\displaystyle \frac{dh}{d \rho} = -\frac{15}{128}(\chi_1f-\chi_2g)V } , \\ 
V^2=1-f^2-g^2-h^2+2fgh ,
\label{ac}
\end{array}
\right.
\end{equation}
whose integral lines in the ($f$, $g$, $h$)-space are arcs of parabolas 
given by the following conditions
\[   \left\{     \begin{array}{l}
h+\frac{1}{2}fg= \mbox{constant}  ,                  \\ 
\chi_1f+\chi_2g = \mbox{constant} , \\ 
1-f^2-g^2-h^2+2fgh \geq 0     .
\end{array}    \right.     \]

We can set a new parameter $u$ as 
\[ 
u=\frac{\chi_2}{\sqrt{\chi_1^2+\chi_2^2}}(f-f_0)-
\frac{\chi_1}{\sqrt{\chi_1^2+\chi_2^2}}(g-g_0) \]
and write $f$ and $g$ as follows
\begin{equation} 
f=f_0+\frac{\chi_2}{\sqrt{\chi_1^2+\chi_2^2}}u , ~~~~~~~~~~~~~~~
g=g_0-\frac{\chi_1}{\sqrt{\chi_1^2+\chi_2^2}}u .
\label{r} \end{equation}

Substituting equations (\ref{r}) into equation (\ref{q}) and both 
equations (\ref{q}) and (\ref{r}) repeatedly into equation (\ref{ac}), we 
can finally reduce the system of equations (\ref{ac}) to quadrature:
\begin{equation}     \left\{     \begin{array}{rl}
h(u)=&{\displaystyle h_0-\frac{(\chi_1f_0-\chi_2g_0)^2}{8\chi_1\chi_2}+\frac{\chi_1\chi_2}
{2(\chi_1^2+\chi_2^2)}\left[u-\frac{\sqrt{\chi_1^2+\chi_2^2}}{2\chi_1\chi_2}
\; (\chi_2g_0-\chi_1f_0)\right]^2 } , \\ { } & { } \\
V^2(u)=&{\displaystyle 1-\frac{(\chi_1f_0+\chi_2g_0)^2}{\chi_1^2+\chi_2^2}+
\frac{1}{5}(2h_0+f_0g_0)^2+ } \\
&{\displaystyle -\left[u-\frac{(\chi_1g_0-\chi_2f_0)}
{\sqrt{\chi_1^2+\chi_2^2}}\right]^2-5\left[h(u)-
\frac{1}{5}(2h_0+f_0g_0)\right]^2 } , \\ {} & {} \\
{\displaystyle \frac{du}{d\rho}}=&{\displaystyle -\frac{15}{64}
\sqrt{\chi_1^2+\chi_2^2} } \; V(u) .
\end{array}        \right.         \label{ad} \end{equation}
The solution, 
\begin{equation}
-\frac{15}{64}\sqrt{\chi_1^2+\chi_2^2}\; (\rho-\rho_0)=\int_0^u 
\frac{dv}{V(v)} ,
\label{ap}   \end{equation}
involves an elliptic integral of the first kind, so that 
$u$ can be expressed in terms of the Jacobian elliptic functions
(\cite{AS}, p.~596).
Let us remark that the oscillation period is inversely proportional to
$(\chi_1^2+\chi_2^2)^{1/2}$ and that the turning points of $u$, i.e.,
the zeros of $V(u)$, are given by the intersection in the $(u, h)$-plane of the parabola
\[                  
h=h_0-\frac{(\chi_1f_0-\chi_2g_0)^2}{8\chi_1\chi_2}+\frac{\chi_1\chi_2}
{2(\chi_1^2+\chi_2^2)}\left[u-\frac{\sqrt{\chi_1^2+\chi_2^2}}{2\chi_1\chi_2}
(\chi_2g_0-\chi_1f_0)\right]^2 
\]
with the ellipse
\[    \begin{array}{rl}
{\displaystyle 
\left[u-\frac{(\chi_1g_0-\chi_2f_0)}{\sqrt{\chi_1^2+\chi_2^2}}\right]^2+}&
{\displaystyle  5\left[h-\frac{1}{5}(2h_0+f_0g_0)\right]^2 = }\\
& = {\displaystyle 1-\frac{(\chi_1f_0+\chi_2g_0)^2}{\chi_1^2+\chi_2^2}+
\frac{1}{5}(2h_0+f_0g_0)^2 } .
\end{array}           \]
We can also notice that the oscillation amplitude depends on $\chi_1$ and 
$\chi_2$ only through their relative weight (e.g., the quantity 
$\chi_1/\chi_2$ or its reciprocal) and is independent from their absolute 
magnitudes.
In fact, the oscillation amplitude is determined only by the right-hand 
side of equation (\ref{ap}), in which $V(v)$ is a homogeneous function of 
$\chi_1$ and $\chi_2$ with degree zero, as an inspection of equation
(\ref{ad}) readily shows.

Let us finally observe that the approximation condition $X_1=X_2$ may in 
practice be replaced by the condition $| X_1-X_2|\rho^{1/2} \ll 
\min\{\chi_1,\chi_2\}$ sufficient for the first term in equation (\ref{n}) 
to be negligible.

\subsubsection{Examples.}

\begin{enumerate}    

\item $\chi_1=\chi_2=\chi \not= 0$.
We have from equation (\ref{ad}) :
\begin{equation}     \left\{    \begin{array}{l}
h=h_0-\frac{1}{4}u_{\ast}^2+\frac{1}{4}(u-u_{\ast})^2 , \\ {} \\
V^2=\frac{5}{16}\left[\frac{8}{5}(\sqrt{\Delta}+a-
\frac{5}{8}u_{\ast}^2)+(u-u_{\ast})^2\right] \left[\frac{8}{5}(\sqrt{\Delta}-
a+\frac{5}{8}u_{\ast}^2)-(u-u_{\ast})^2 \right] ,
\end{array}     \label{j}        \right.     \end{equation}
where $u_{\ast}=(g_0-f_0)/2^{1/2}$,
$a=1+(3h_0-f_0g_0)/2$, $\Delta=a^2+5 V_0^2/4$.
Equation (\ref{ap}) can be integrated now by direct application of one 
of the formulae on p.~596 of (\cite{AS}), the selection of which depends
on the signs of the quantities $a-5 u_{\ast}^2/8 \pm \Delta^{1/2}$.
For example, if we suppose that $u_{\ast}=0$, that is $f_0=g_0$, then 
$a+\Delta^{1/2} \geq 0$ and $a-\Delta^{1/2} \leq 0$.
The solution is then 
\begin{equation}
u(\rho)=-\mbox{sign}(V_0)\sqrt{\frac{4(\Delta-a^2)}{5\sqrt{\Delta}}}
\; \; \mbox{sd}\left(\left.\frac{15}{64}\chi\sqrt[4]{4\Delta}(\rho-\rho_0) 
\right| \frac{\sqrt{\Delta}-a}{2\sqrt{\Delta}}\right) , \label{ca}
\end{equation}
where ``sd'' is the well-known Jacobian elliptic function,
equivalent to the function ``cn'' modulo a translation and a rescaling:
$\mbox{sd}(v|y)=\mbox{cn}(v-K(y)|y)/(1-y)^{1/2}$ [see eq.\ (16.8.2) 
in \cite{AS}; from now on $K(y)$ is the complete elliptic integral 
of the first kind with parameter $y$, as defined by eq.\ (17.3.1) in 
the last reference].
As $\rho$ varies, $u(\rho)$ oscillates with a period
\begin{equation}
\lambda = \frac{256 \; K\! \left( (\sqrt{\Delta}-a)/(2\sqrt{\Delta})
\right)}{15\sqrt[4]{4\Delta}\; \chi} .    \label{m}
\end{equation}

The analytic equations (\ref{j}), (\ref{ca}), and (\ref{m}) show 
the dependence on the initial conditions $f_0$, $g_0$, $h_0$ explicitly.
Let us notice that the amplitude of oscillations does not depend on 
$\chi$, in this case.
A plot of $f(\rho)$ and $V(\rho)$, together 
with the numerical solution of equations (\ref{w})--(\ref{y}) 
for comparison, is shown in Figure 6 for the choice of
$\rho_0=75$, $f_0=g_0=0.25$, $h_0=-0.5$, $V_0=-0.75$, $\chi_1=1$, 
$\chi_2=1$, corresponding to Figure 12 of (\cite{Aea}).
Here we have $a=7/32\approx 0.22$, $\Delta=769/1024\approx 0.75$.
In this case we notice that, close to the last stable orbit, the agreement 
between analytical and numerical calculations becomes slightly less 
satisfactory. 

\item $\chi_1/\chi_2 \rightarrow 0$.   \label{an}
We have from equations (\ref{r}) and (\ref{q}) that $f=f_0+u$, $g=g_0$ and 
$h=h_0-g_0 u/2$.
Thus, equation (\ref{y}) yields
\[
V^2(u)=a^2-b^2(u-u_{\ast})^2 ,
\]
where $b^2=1+5 g_0^2/4$, 
$u_{\ast}=(3g_0h_0-2f_0-f_0g_0^2)/(2b^2)$, 
$a^2=V_0^2+b^2u_{\ast}^2$.
Therefore the equation $du/d\rho= -15 \chi_2V(u)/64$
is easily integrated and yields
\begin{equation}
u(\rho)=u_{\ast}-(\mbox{sign}V_0)\frac{a}{b}\sin\left[\frac{15}{64}b\chi_2
(\rho-\rho_0)+(\mbox{sign}V_0)\varphi\right] ,
\label{ae}    \end{equation}
where $\sin\varphi = bu_{\ast}/a$.
As $\rho$ varies $u$ oscillates with a period given by the following 
expression 
\begin{equation}  
\lambda=\frac{128\pi}{15\sqrt{1+5 g_0^2/4}\;\chi_2} ,
\label{af}    \end{equation}
in perfect agreement\footnote{There are a few minor typographic errors 
in Appendix B of (\cite{CF}).
Equations (B14), (B20) and (B25) should read 
$h_4 = -15(s_2\alpha_{2,i}+L\delta)/(128\mu)$,
$\kappa^2=1-\alpha_{2,i}^2-(1-\alpha_{2,i}^2)\alpha_{-,i}^2/\nu_0^2$ 
and $\nu_0^2=\left[ 225/4096+1125 \alpha_{2,i}^2/16384 \right]
s_2^2/\mu^2$ respectively.} with formula (B25) of (\cite{CF}).

\item $f_0=g_0=0$.
We have from equation (\ref{ad}) that the following relations hold:
\begin{equation}
\left\{    \begin{array}{l}
{\displaystyle h=h_0+\frac{\Gamma}{2}u^2  } , \\ {} \\
{\displaystyle V^2=\frac{5}{4}\Gamma^2(u^2+a^2)(b^2-u^2) },
\end{array}             \right.  \label{cb}   \end{equation}
where we have defined the following constants: 
$\Gamma = \chi_1\chi_2/(\chi_1^2+\chi_2^2)$,
$a^2= \{ 2[1+6h_0\Gamma+(5+4h_0^2)\Gamma^2]^{1/2}+2+6h_0\Gamma\}/
(5\Gamma^2)$ and 
$b^2= \{ 2[1+6h_0\Gamma+(5+4h_0^2)\Gamma^2]^{1/2}-2-6h_0\Gamma\}/
(5\Gamma^2)$.
The solution is
\begin{equation}
u(\rho)=\frac{-\mbox{sign}(V_0)ab}{\sqrt{a^2+b^2}}
\; \mbox{sd}\left(\frac{15}{128}\sqrt{5(\chi_1^2+\chi_2^2)\Gamma^2(a^2+b^2)}
(\rho-\rho_0) \left| \frac{b^2}{a^2+b^2}\right. \right) .
\label{ag}        \end{equation}
As $\rho$ varies $u$ oscillates with a period
\begin{equation}
\lambda = \frac{256 \; K\! \left(b^2/(a^2+b^2)\right)}
{15\sqrt{\chi_1^2+\chi_2^2}\; \sqrt[4]{1+6h_0\Gamma+(5+4h_0^2)\Gamma^2}} .
\label{ab}        \end{equation}

The analytic equations (\ref{cb}), (\ref{ag}), and (\ref{ab}) show 
the dependence on $\chi_1$ and $\chi_2$ explicitly. 
Both the oscillation amplitude and period depend on $\chi_1$ and 
$\chi_2$ in this case, but the amplitude depends only on $\Gamma$ and 
not on $(\chi_1^2+\chi_2^2)^{1/2}$, as previously noticed.
Let us remark that in the limit $\chi_1 \ll \chi_2$ we find for 
equations (\ref{ag}) and (\ref{ab}) the same expressions given 
in the example (\ref{an}) by equations (\ref{ae}) and (\ref{af}) 
with $f_0=g_0=0$, as is easily seen when we notice that, if 
$\chi_1 \ll \chi_2$, then $\Gamma \sim \chi_1/\chi_2$, 
$a^2 \sim 4\chi_2^2/(5\chi_1^2)$, $b^2 \sim 1-h_0^2$, and that 
$\lim_{y\rightarrow 0}\mbox{sd}(v|y)=  \sin v$.

We plot in Figure 7 the analytical solutions ({\em solid line}) 
and the numerical solutions ({\em dots})
for $f(\rho)$ and $V(\rho)$, with the chosen values 
$\chi_1=10^{-3}$, $\chi_2=1$, $f_0=g_0=h_0=0$ and 
$\rho_0 =250$.
As we see, the agreement between the analytical and the numerical 
curves is striking.

\end{enumerate}

\subsection{Very different masses at large separations.} \label{v}

By differentiating equation (\ref{y}) twice with respect to $\rho$ and making 
repeated use of equations (\ref{w})--(\ref{n}), we obtain
$d^2V/d\rho^2 =F(f,g,h,\rho)$ 
where $F$ is an algebraic function, easily calculated.
If $| X_1-X_2 | \rho^{1/2} \gg 1$, then the leading term in F is much greater 
than all others and the equation for $V$ is approximated by the 
following one
\[
\frac{d^2V}{d\rho^2}=-\left(\frac{15}{128} \frac{X_1-X_2}{\eta}\right)^2\rho
V(\rho) ,
\]
whose solution is
\begin{equation}
V(\rho)=C_1 \mbox{ Ai }(-\Omega^{2/3}\rho)+C_2 \mbox{ Bi }(-\Omega^{2/3}\rho) ,
\label{s} \end{equation}
where Ai and Bi are the Airy functions
(fully described in Appendix b of \cite{LL3} and \S 10.4 of 
\cite{AS}),
$\Omega= 15 | X_1-X_2|/(128\eta)$, $C_1$ and $C_2$ are integration
constants such that $V(\rho_0)=V_0$ and $\left[ dV/d\rho
\right]_{\rho_0}=V'_0$, namely,
\[
\left\{    \begin{array}{l}
{\displaystyle C_1=\frac{V_0 \mbox{ Bi}'\;(-\Omega^{2/3}\rho_0)+V'_0 \;
\Omega^{-2/3} \mbox{ Bi }(-\Omega^{2/3}\rho_0)}{\mbox{ Ai }
(-\Omega^{2/3}\rho_0) \mbox{ Bi}'\;(-\Omega^{2/3}\rho_0)-\mbox{ Ai}'\;
(-\Omega^{2/3}\rho_0) \mbox{ Bi }(-\Omega^{2/3}\rho_0)} } , \\  {} \\
{\displaystyle C_2=\frac{V_0 \mbox{ Ai}'\;(-\Omega^{2/3}\rho_0)+V'_0 \;
\Omega^{-2/3} \mbox{ Ai }(-\Omega^{2/3}\rho_0)}{\mbox{ Ai}'\;
(-\Omega^{2/3}\rho_0) \mbox{ Bi }(-\Omega^{2/3}\rho_0)-\mbox{ Ai }
(-\Omega^{2/3}\rho_0) \mbox{ Bi}'\;(-\Omega^{2/3}\rho_0)} } .
           \end{array}         \right.
\]

We then get, by truncating equations (\ref{w})--(\ref{n}) to the lowest order terms 
in $\rho$, the following approximate solutions:
\begin{equation}   \left\{   \begin{array}{l}
{\displaystyle f(\rho) = f_0 -\frac{15}{128}\frac{\chi_2}{X_1}
\int_{\rho_0}^{\rho} V(y) \, dy } , \\ {} \\
{\displaystyle g(\rho) = g_0 +\frac{15}{128}\frac{\chi_1}{X_2}
\int_{\rho_0}^{\rho} V(y) \, dy } , \\  {} \\
{\displaystyle h(\rho) = h_0 +\frac{15}{128}\frac{X_1-X_2}{\eta}
\int_{\rho_0}^{\rho} y^{1/2}V(y) \, dy } ,
\end{array}  \right.   \label{t}   \end{equation}
where $V(\rho)$ is given by equation (\ref{s}).

For the evaluation of equations (\ref{t}) we need to know 
the functions $\int^y \mbox{Ai}\;(-y)\,dy$ and $\int^y 
\mbox{Bi}\;(-y)\,dy$, 
which are tabulated in Table~10.12 of (\cite{AS}), or asymptotic 
expansions for them (valid if both $\rho$ and $\rho_0$ are much greater 
than unity), which we adapted from equations (10.4.83) and (10.4.85) of 
(\cite{AS}) and give here for reference:
\[    \left\{ \begin{array}{l}
{\displaystyle \int_{\rho_0}^{\rho} \mbox{ Ai }(-\Omega^{2/3}y)
\, dy \sim - \pi^{-1/2}\Omega^{-7/6}\left[
y^{-3/4}\cos\left(\frac{2}{3}\Omega
y^{3/2}+\frac{\pi}{4}\right)\right]_{\rho_0}^{\rho} } , \\ {} \\
{\displaystyle \int_{\rho_0}^{\rho} \mbox{ Bi }(-\Omega^{2/3}y)
\, dy \sim + \pi^{-1/2}\Omega^{-7/6}\left[
y^{-3/4}\sin\left(\frac{2}{3}\Omega
y^{3/2}+\frac{\pi}{4}\right)\right]_{\rho_0}^{\rho} } .
\end{array}      \right.      \]
We also need the following formulae, which we worked out with the help of 
equations (10.4.60) and (10.4.64) in (\cite{AS}):
\[    \left\{ \begin{array}{l}
{\displaystyle \int_{\rho_0}^{\rho} y^{1/2}\mbox{ Ai }(-\Omega^{2/3}y)
\, dy \sim - \pi^{-1/2}\Omega^{-7/6}\left[
y^{-1/4}\cos\left(\frac{2}{3}\Omega
y^{3/2}+\frac{\pi}{4}\right)\right]_{\rho_0}^{\rho} } , \\ {} \\
{\displaystyle \int_{\rho_0}^{\rho}y^{1/2}\mbox{ Bi }(-\Omega^{2/3}y)
\, dy \sim + \pi^{-1/2}\Omega^{-7/6}\left[
y^{-1/4}\sin\left(\frac{2}{3}\Omega
y^{3/2}+\frac{\pi}{4}\right)\right]_{\rho_0}^{\rho} } .
\end{array}      \right.      \]

If $X_1\not= X_2$, the condition $| X_1-X_2 | \rho^{1/2} \gg 1$ for the 
validity of the approximation can always apply at sufficiently large 
values of $\rho$, but the agreement becomes increasingly poorer 
as $\rho$ decreases.
In fact, at short separations the solution given by equations 
(\ref{t}) fails to predict both the period and the amplitude of 
oscillations to a sufficient degree of accuracy. 
Nevertheless this 
solution is worth being considered because it is very 
simple and approximately describes the behavior of the system in 
a large variety of cases.

\newpage

\section*{Figure captions.}

\begin{itemize}

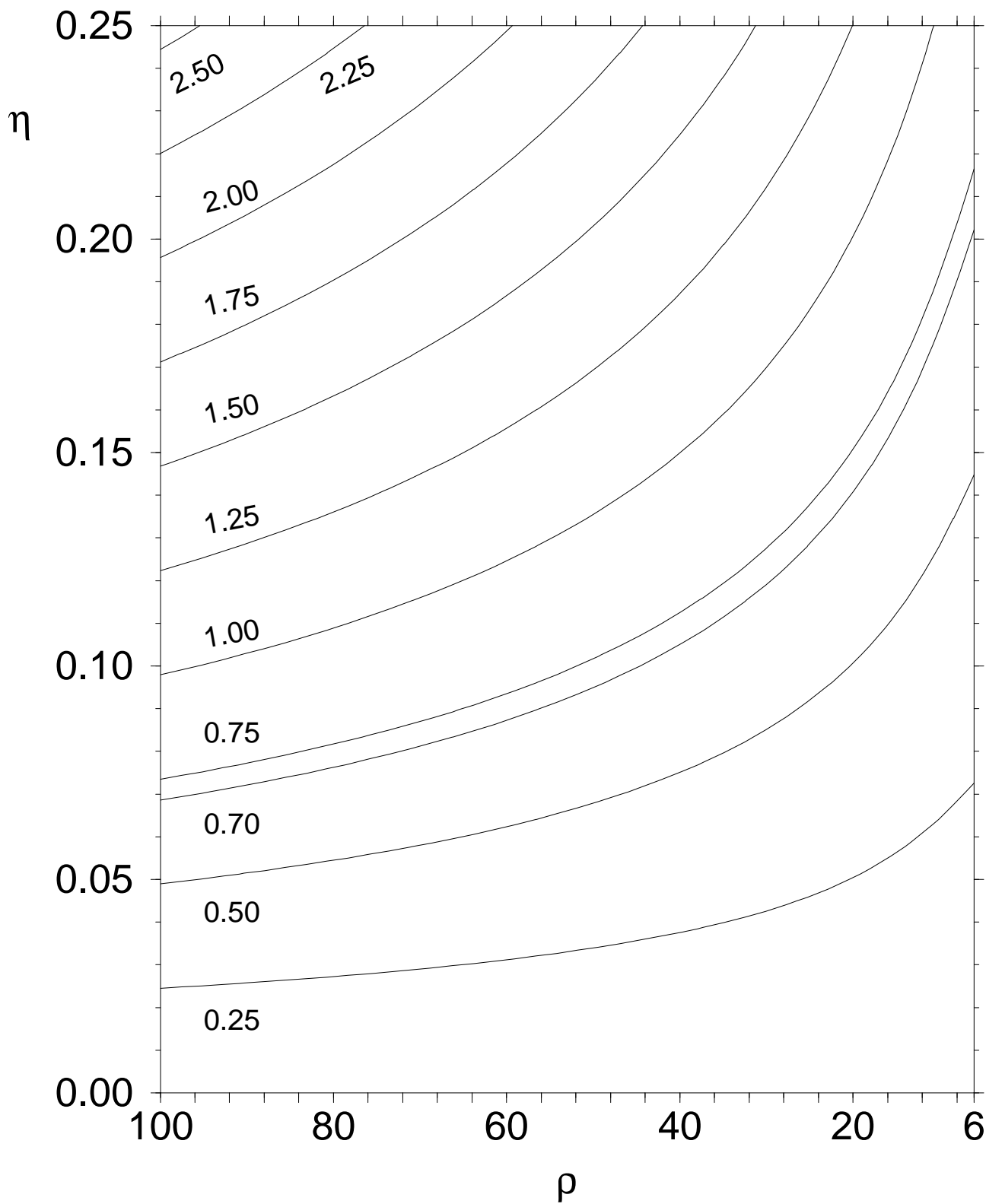
\item[Fig.\ 1.] Contour lines $cJ/(GM^2)=0.25, 0.50, \ldots, 2.50$ 
in the ($\rho, \eta$)-plane.

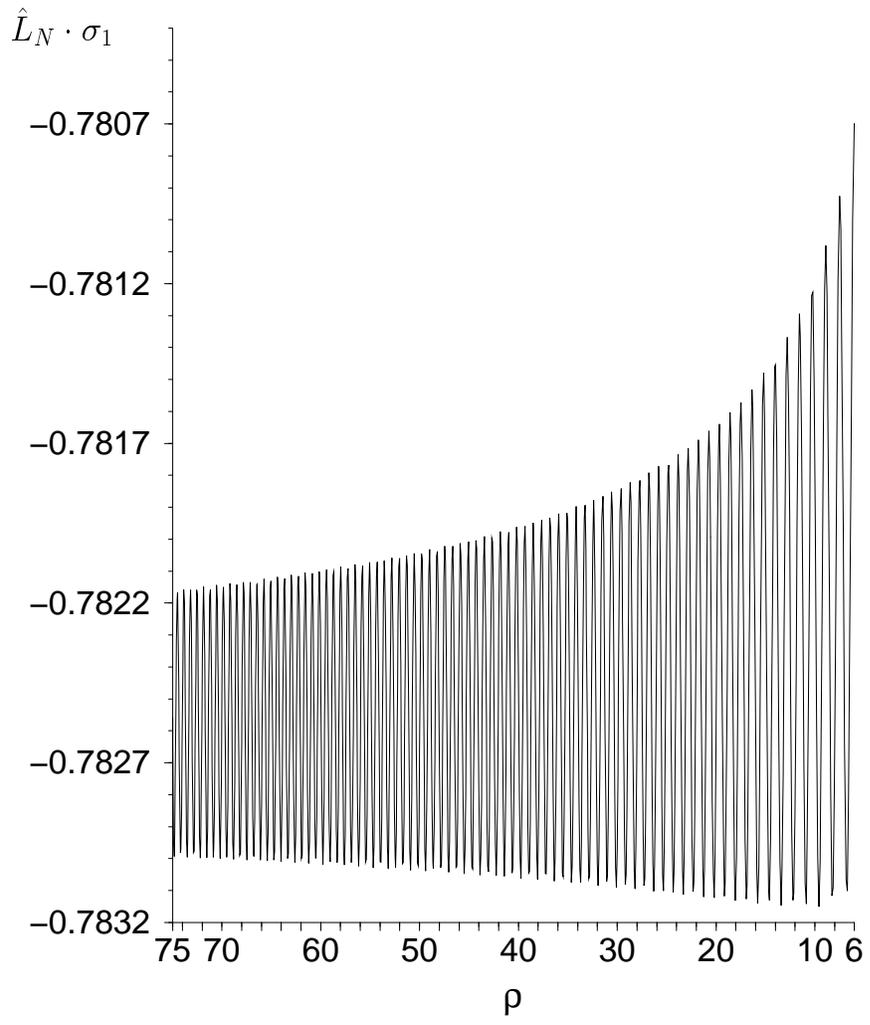
\item[Fig.\ 2.] Evolution of
$\hat{{\bf L}}_N\cdot\mbox{{\boldmath $\sigma$}}_1$ 
from $\rho=75$ to $\rho=6$ for a system with $m_2=0.13m_1$ with 
the following choice of initial values:
$(\hat{{\bf L}}_N\cdot\mbox{{\boldmath $\sigma$}}_1)_0=-0.78256$, 
$(\hat{{\bf L}}_N\cdot\mbox{{\boldmath $\sigma$}}_2)_0=-1.1177 \times 
10^{-4}$, $(\mbox{{\boldmath $\sigma$}}_1\cdot\mbox{{\boldmath 
$\sigma$}}_2)_0=-8.7585 \times 10^{-5}$,
$(\hat{{\bf L}}_N\cdot \mbox{{\boldmath $\sigma$}}_1\times
\mbox{{\boldmath $\sigma$}}_2)_0=-3.6197 \times 10^{-4}$.

\item[Fig.\ 3.] Behavior of the Kerr parameter as function of $\rho$ 
when $\chi_1=\chi_2=1$ ({\em solid lines}) and
\begin{enumerate}
\item[(a)] $m_1=m_2$; $f_0=g_0=0.9,~ h_0=0.63$ at $\rho_0 = 100$.
\item[(b)] $m_2=0.1m_1$; $f_0=-g_0=0.9,~ h_0=-0.7$ at $\rho_0 = 100$.
\item[(c)] $m_2=0.1m_1$; $f_0=-0.42555, ~g_0=-0.59746,~ h_0=0.23651$ at 
$\rho_0 = 180$.
\item[(d)] $m_2=0.13m_1$; $f_0=-0.999247282, ~g_0=-8.44485 \times 10^{-3},
~ h_0=-8.45 \times 10^{-3}$ at $\rho_0 = 75$, which is the same choice as 
for Fig.\ 2.
\end{enumerate}
The corresponding spinless cases are shown for comparison ({\em dashed 
lines}).

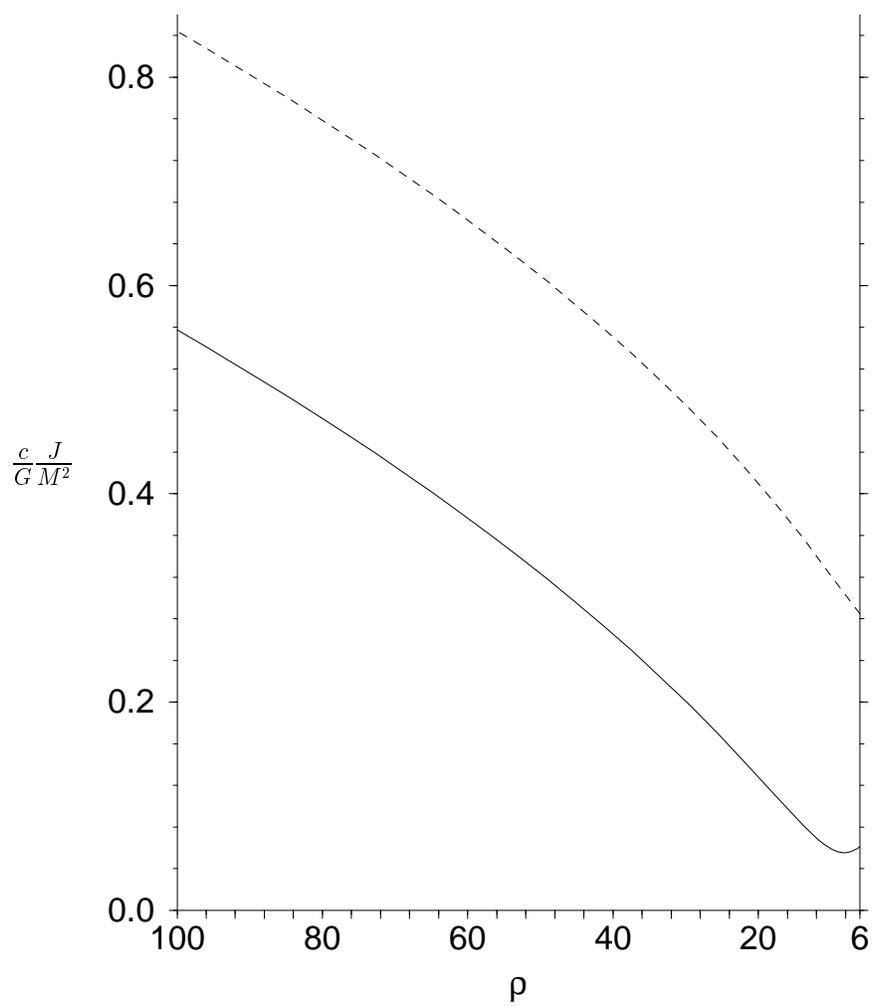
\item[Fig.\ 4.] Decrease of the Kerr parameter almost to zero at 
the end of the inspiral phase is obtained only for particular choices 
of parameters and initial conditions. 
In this case we have $m_2=10 m_1$, $\chi_1=\chi_2=0.351$ and 
$f_0=-g_0=-h_0=1$ at $\rho_0=100$.

\item[Fig.\ 5.] (a) Probability density function of $cJ/(GM^2)$ at $\rho=6$ 
computed by the Monte Carlo method. 
Assumptions for the probability distributions of parameters and 
relative orientations of angular momenta are explained in text.
A normal distribution with same mean and standard deviation is 
plotted for comparison.
(b) Comparison of the same curve as in (a) with the probability density 
function of $cJ/(GM^2)$ at $\rho=6$ for the spinless case ({\em dashed 
line}), numerically computed. Notice that the figure scale is different from 
(a) to (b).

\item[Fig.\ 6.] Plots of (a) $f = \hat{{\bf L}}_N\cdot\hat{{\bf s}}_1$, 
(b) $V = \hat{{\bf L}}_N\cdot\hat{{\bf s}}_1\times\hat{{\bf s}}_2$
versus separation $\rho$, when $\rho_0=75$; $f_0=g_0=0.25$,
$h_0=-0.5$, $V_0=-0.75$; $\chi_1=\chi_2=1$ and $m_1=m_2$. 
Solid lines show the analytical solutions, dots show the numerical 
solutions to equations (\ref{w})--(\ref{y}).

\item[Fig.\ 7.] Plots of (a) $f = \hat{{\bf L}}_N\cdot\hat{{\bf s}}_1$, 
(b) $V = \hat{{\bf L}}_N\cdot\hat{{\bf s}}_1\times\hat{{\bf s}}_2$
versus separation $\rho$, when $\rho_0=250$; $f_0=g_0=h_0=0$;
$V_0=-1$; $\chi_1=10^{-3}$, $\chi_2=1$ and $m_1=m_2$. 
Solid lines show the analytical solutions, dots show the numerical 
solutions to equations (\ref{w})--(\ref{y}).

\end{itemize}

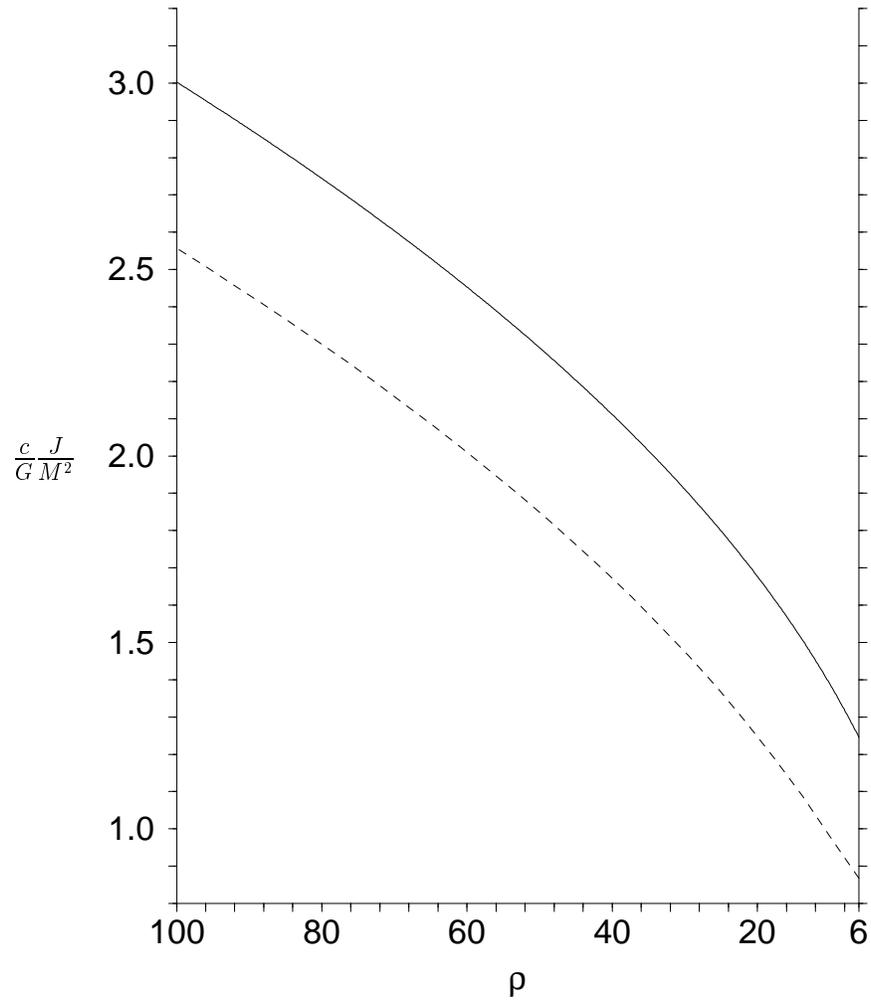
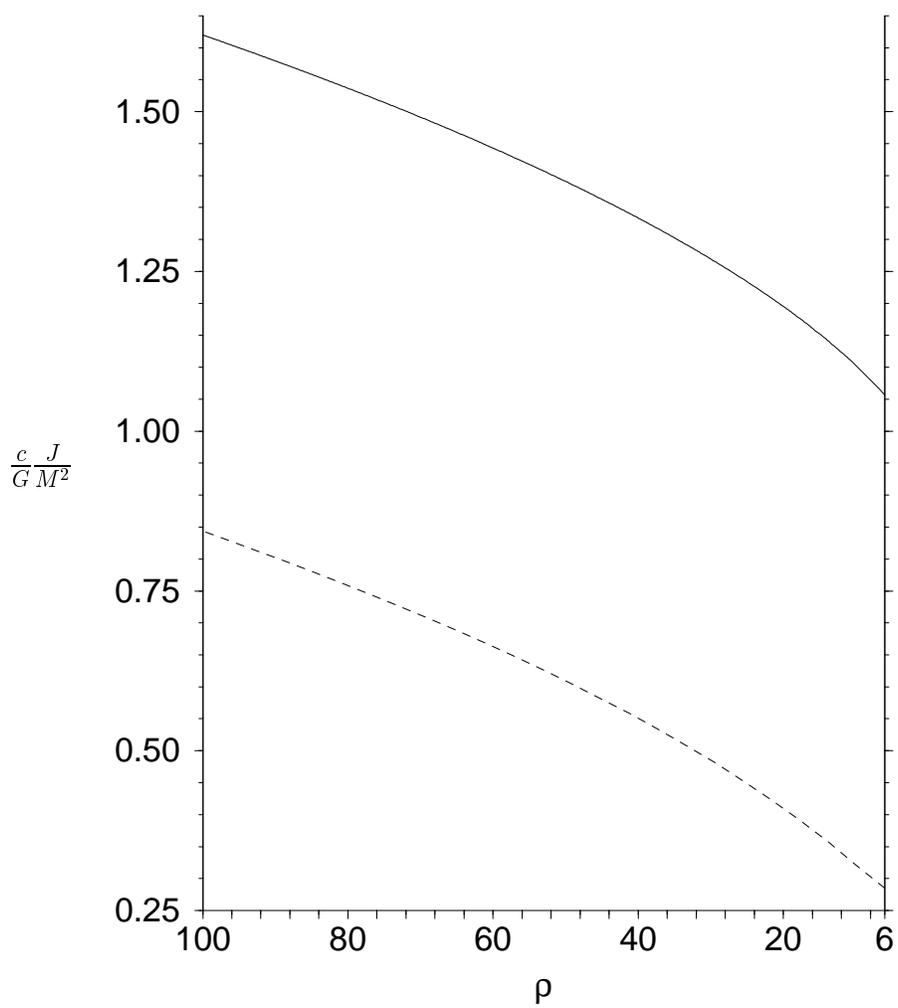
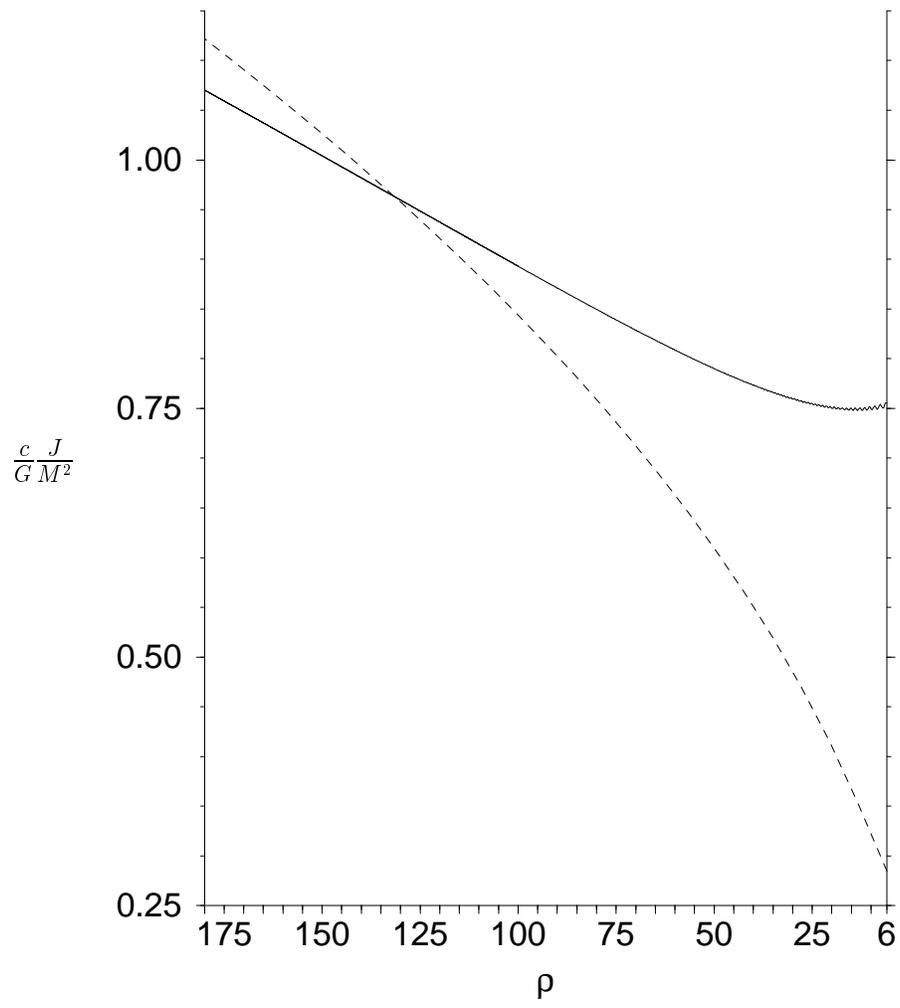
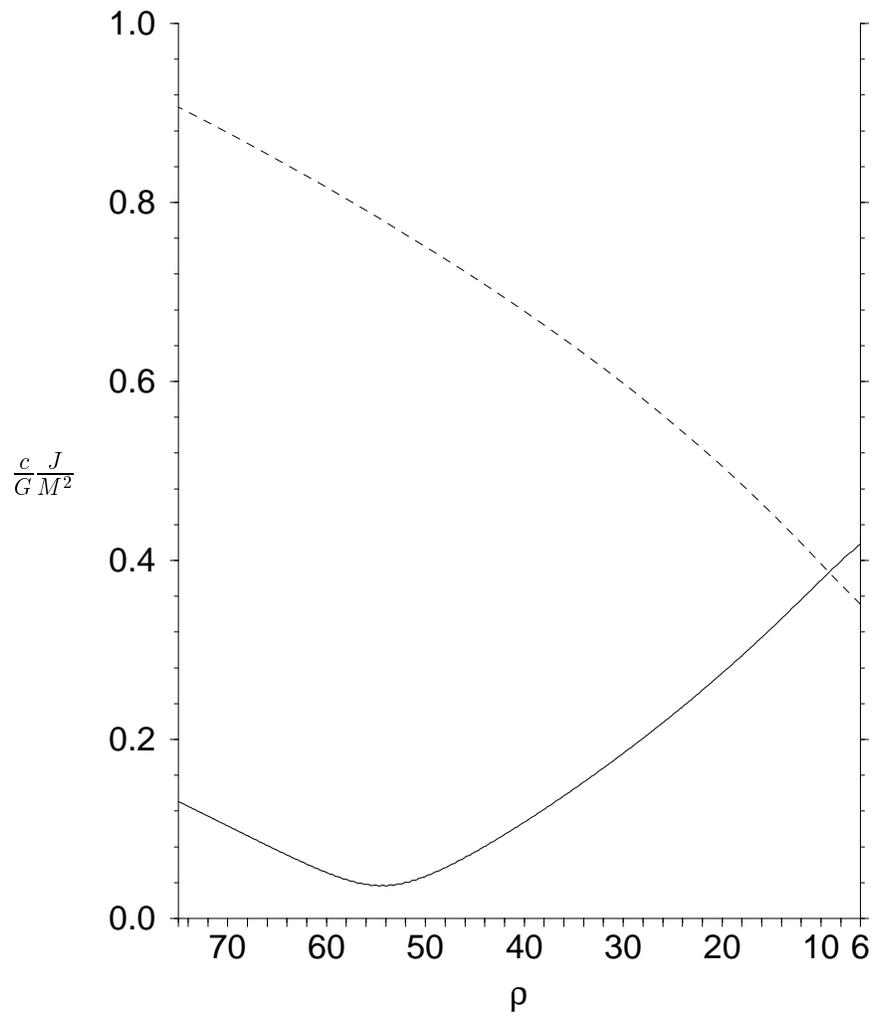
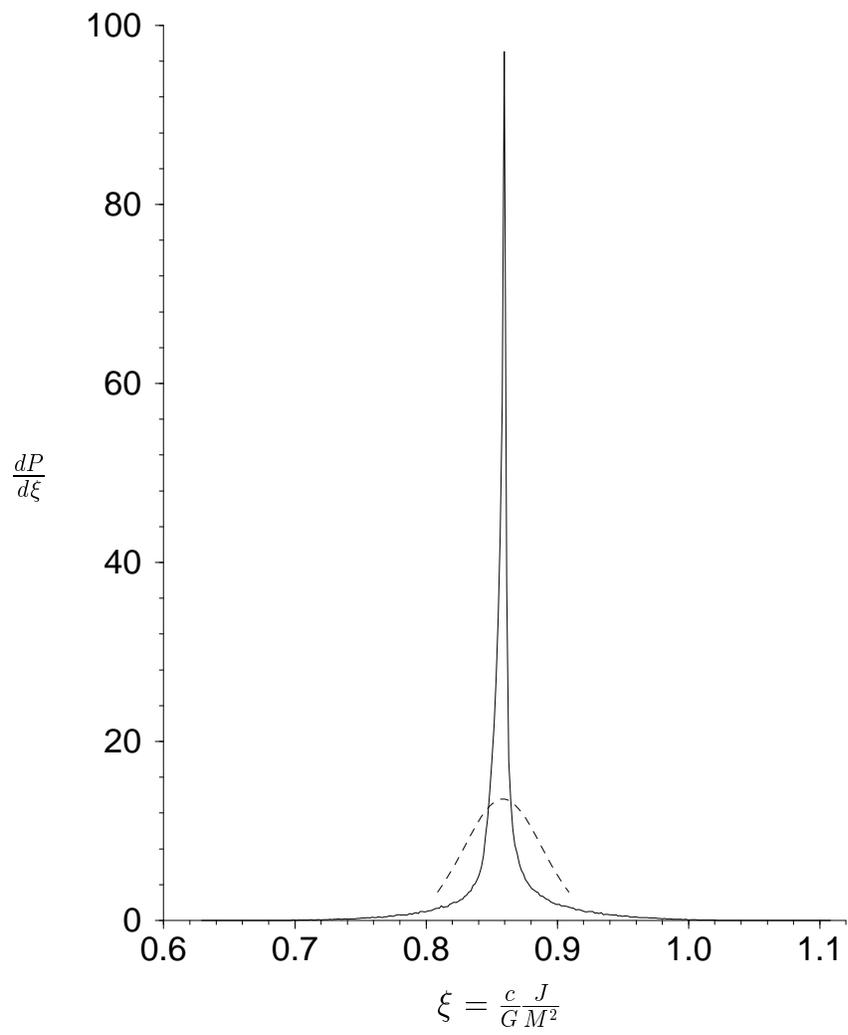
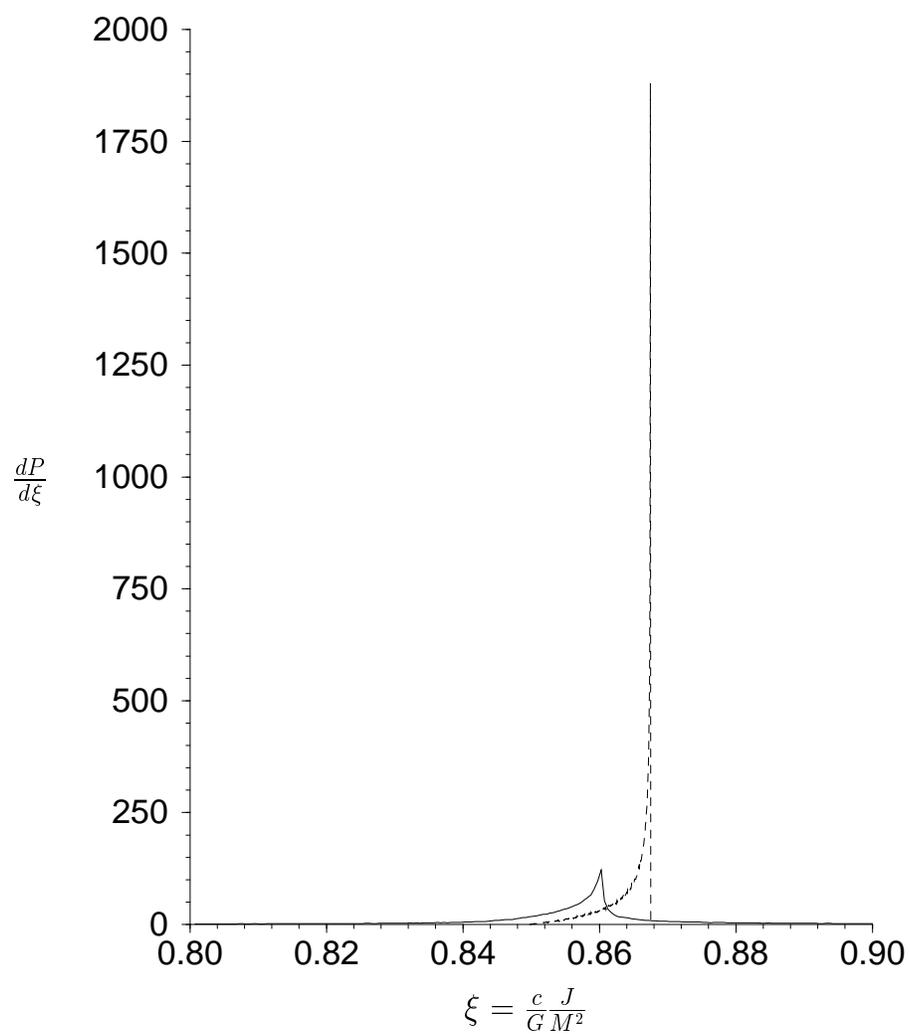
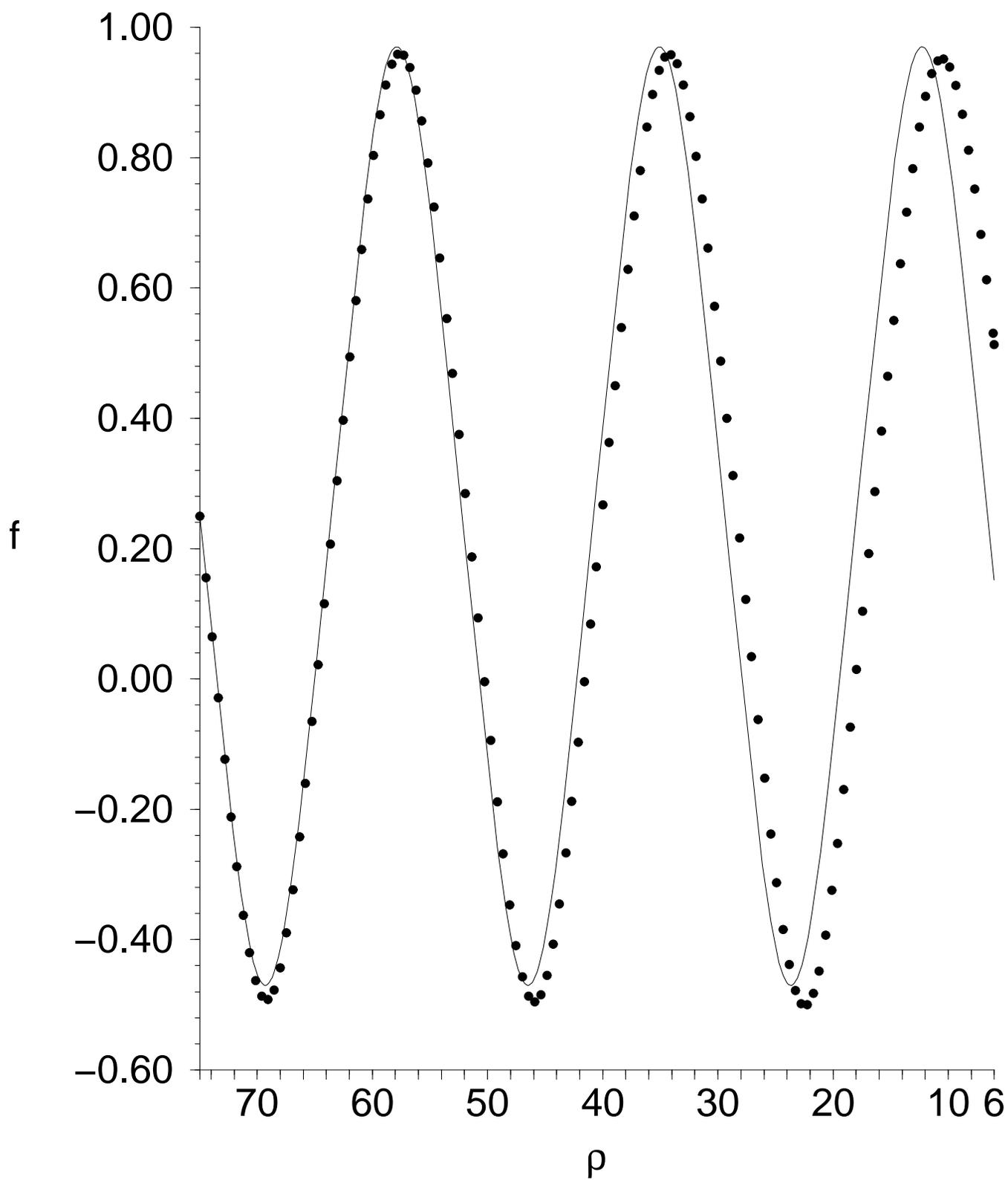
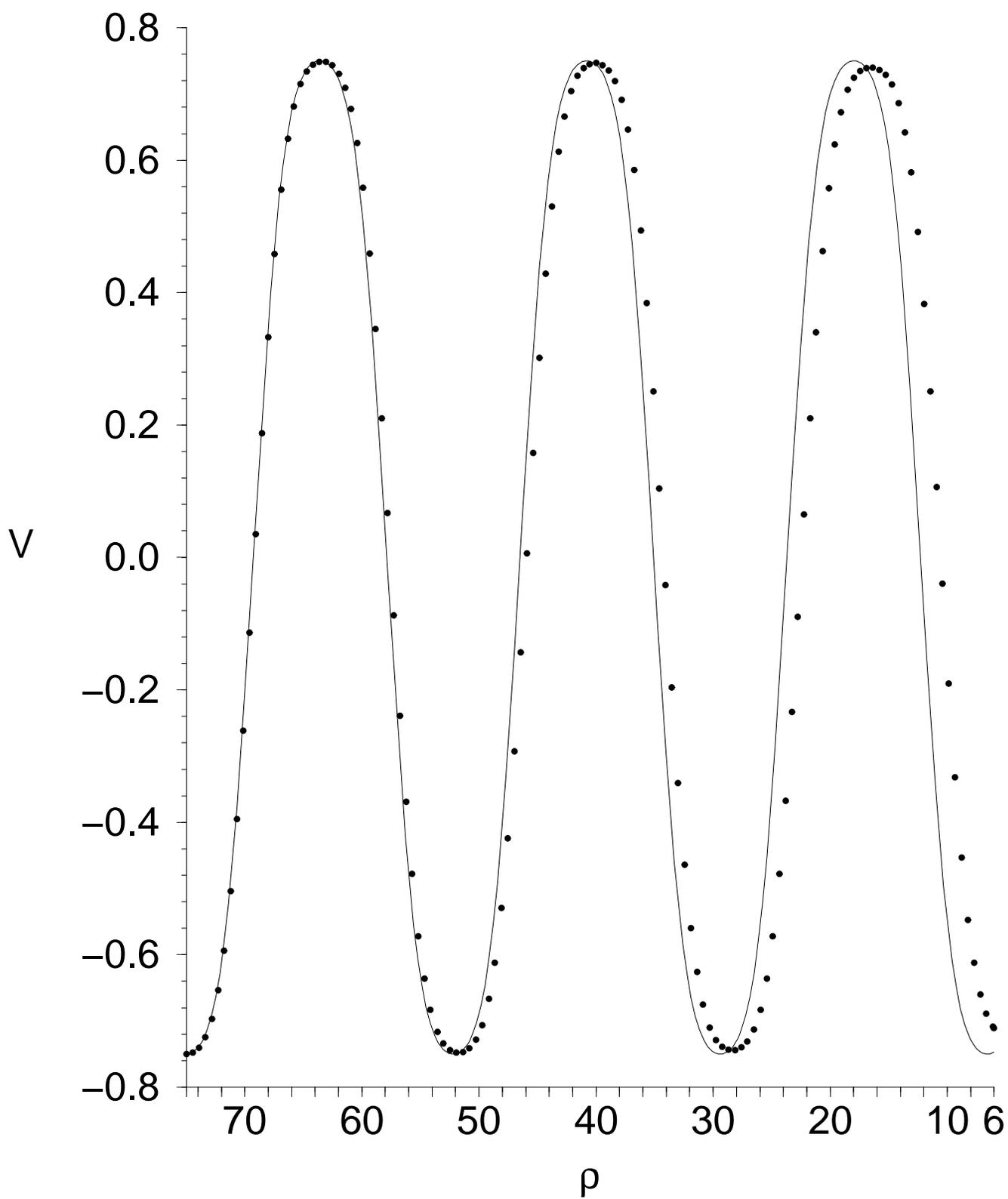
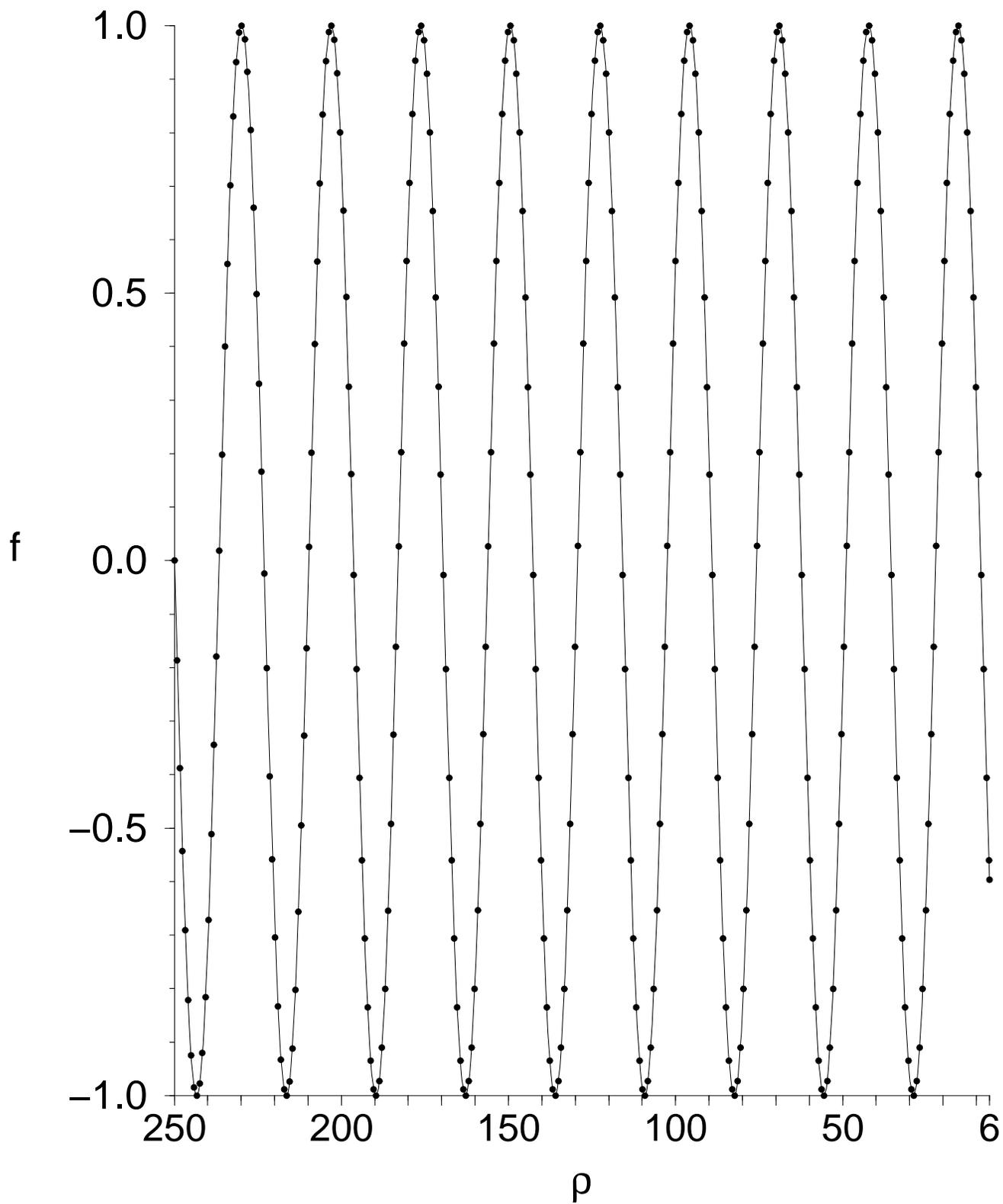
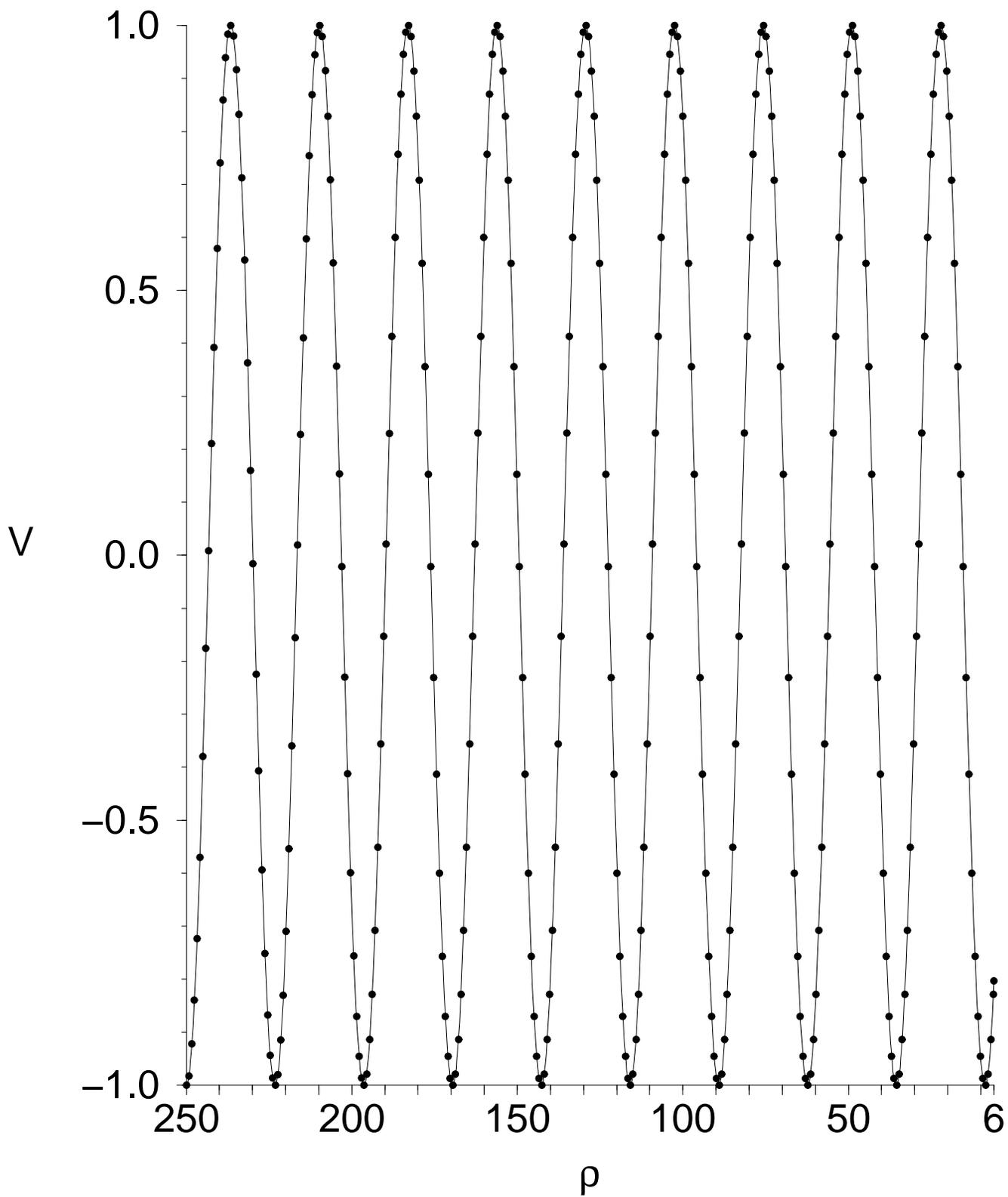
\end{document}